\documentclass[aps,prb,showpacs]{revtex4}
\usepackage{amsmath}
\usepackage{amssymb}
\usepackage{epsfig}

\begin{document}
\title{Josephson junctions chains with long-range interactions,
 phase slip proliferation versus Kosterlitz-Thouless transition}
\author{P. Devillard$^{1,2}$}
\affiliation{$^1$ Centre de Physique Th\'eorique de Marseille
        CPT, case 907, 13288 Marseille Cedex 9 France}
\affiliation{$^2$ Universit\'e de Provence, 3, Place Victor Hugo, 
13331 Marseille cedex 03, France}

\begin{abstract}
The role of long range badly screened Coulomb interactions in a one-dimensional
 chain of Josephson junctions is studied. Correlation functions for the 
phase correlator are obtained as a function of the Josephson coupling energy,
 the short range part of Coulomb repulsion and its long range component.
 Though quasi-long range order is no longer possible and the usual Kosterlitz-Thouless
 transition no longer exists, there are remnants of it. 
 As an application, 
we calculate the $I-V$ curves for Andreev reflexion when a normal metal
 is placed in contact with the chain. Formally, there is always an offset voltage $V_0$
 below which no current can flow, however, in some regimes $V_0$ can be negligible. 
Contrary to what happens without long-range 
interactions, the Andreev current, as a function of applied 
voltage, increases faster than any power law. Signatures of long range interactions 
and phase
 slips appear in the $I-V$ curves. Possible application
 for quasi one-dimensional thin superconducting wires is outlined.
 \end{abstract}
\maketitle

\section{Introduction}
\label{sec:introduction}

Josephson junction arrays (JJA) have been widely studied, both theoretically and
 experimentally\cite{Fazioreview}. 
They show a number of phenomena,
 vortex physics\cite{BerezinskiiKosterlitzThouless}, quantum phase transitions. They are
 also a paradigm; a large variety of physical systems can be modelled using 
Josephson junctions arrays, for example, one-dimensional thin
 superconducting wires\cite{Zaikin,HalperinDemler}
 and granular superconductors. Nowadays, they are used to realize
 protected qbits\cite{Doucot} or for 
metrological applications, for establishing current standards 
through frequency measurements of Shapiro steps\cite{GuichardBuisson}.

Here, we shall restrict ourselves to the 1-dimensional case. Furthermore, we shall neglect dissipation, which is crucial in some quantum wires. No coupling to a dissipative environment
 is considered. For normal regular junction chains, the essential
 physics is the competition between
 Josephson coupling and local Coulomb interactions. In one dimension, 
true long range order can not exist, the correlations of
 the superconducting order parameter
 decay at best as a power law. When local Coulomb interactions are increased, the
 power law is modified, and eventually a
 Berezinskii-Kosterlitz-Thouless (BKT) transition occurs. 
Beyond the transition, when Coulomb
 interaction dominates, the decay of correlations becomes exponential. This is due to
 the unbinding of phase slips, whose existence have been shown directly
 in recent experiments  \cite{GuichardBuisson}.

Topological defects have always played a central role, from BKT transitions in magnets,
 roughening of interfaces, disordered Luttinger liquids and some stripe models 
for high-$T_c$ superconductors. 

We would like in this paper to study the role of long range Coulomb interactions, 
hereafter denoted as LRI, 
which have not been taken into account so far in one-dimensional JJ chains. 
One motivation is that, in one-dimensional clean superconducting wires,
 the long range part of the Coulomb interactions is poorly screened. For example,
 the plasma mode frequency goes to zero for small frequency\cite{MooijSchon}. 
This has motivated some recent studies\cite{LobosGiamarchi}.
 Clearly, in these systems, fluctuations of the phase of the order
 parameter have to be taken into account.
 The Hamiltonian looks very much like the one of a 
one-dimensional Josephson array, with a long range Coulomb term in addition. 

Long range interactions, hereafter denoted by LRI, were studied a long time ago
 at the spin-wave level for electronic systems\cite{Schulz,Giamarchireview}.
 They also appear naturally in modelling
 underdoped high-$T_c$ superconductors, where phase fluctuations of the order
 parameter are important\cite{dePalo,Randeria}. Influence
 of LRI have also been studied recently\cite{Efetov} theoretically 
in two and three dimensions. Our goal is to obtain the phase correlator of the 
superconducting order parameter as a function of time and space. This quantity
 governs numerous physical properties. In this paper, we shall give only one application. 
Namely, we calculate the average Andreev current from a normal metal to a JJ chain. 
Because LRI are normally screened in fabricated JJ chains, the model discussed here 
 is not meant to be a realistic model for a JJ chain but can be relevant
 to thin 1-dimensional superconducting wires. However, calculations
 developped in this paper can be adapted to take into account the screening
 of the Coulomb interaction between islands and make predictions about 
 Andreev current in real JJ chains.    

After the introduction, the body of the paper is organized in four parts. 

In part two, we first describe the model of JJ chain we use, and its
 connection to high-$T_c$ materials. It is convenient to go to
 imaginary time. When the Josephson coupling dominates over the Coulomb
repulsion, if phase slips are neglected, the modes are spin waves. The long range 
interactions cause the decay of correlations to change
 from those of a Luttinger liquid to a more pronounced decay\cite{Falci}.

 In the next section, we take into account phase slips. The Hamiltonian decouples 
into a part describing spin-waves (long wavelength excitations) and another part 
describing phase slips (rapid variations of the phase). In the last sections, we discuss
 the interaction between phase slips. For distances much smaller than a 
characteristic distance, they interact logarithmically. Beyond this distance,
 the interaction behaves only as the square root of the logarithm
 of the distance\cite{MPAFisher}. 

In part three, we examine how the KT real space renormalization group (RG) analysis
 is modified by LRI. There, as soon as LRI are present, strictly speaking,
 no  quasi long range 
order is possible anymore.
 However, the initial long range interactions can be very weak.
 If we iterate the RG flow, it will take some time for LRI to alter the behavior 
of the usual KT flow. The LRI will show up only after a certain crossover
 ``time'', which we calculate
 and corresponds to some temperature $T^*$. On the other hand, RG has to be stopped
 at some cutoff ``time'', $\hbar/ k_B T$. Therefore, in order to see the influence of LRI,
 the temperature must be lower than $T^*$.  The dependence of $T^*$ on 
the values of parameters such as the Coulomb energy, Josephson coupling and
 strength of LRI is explicited. 

Part four is devoted to the calculation of the phase correlation function.
 Analytical continuation is required to obtain the correlator in real
 time. This part is rather technical; some aspects of the calculations 
are relegated to appendices. The main result is the expression of the 
phase correlator, which differs from the usual Ornstein-Zernike expression. 

In the last part, we discuss one physical situation.
 The Andreev current between a metal and a JJ chain, as a function of applied
 voltage is calculated. We compare the curves obtained in three cases. 
The first case corresponds to only on-site Coulomb interaction. In the second 
case, a Coulomb interaction between nearest-neighbor islands is added.
 This model is often used in the literature. Finally, in the third case,
 LRI are added. For relatively low voltages, the Andreev current remains always
 smaller with LRI. We focuse on the case where, without LRI, we would be in the disordered
 phase. There is a threshold voltage $V_0$ below which no Andreev current can flow.
 This remains true in the presence of LRI. However, the current versus voltage 
($I-V$) curves are different from the case without LRI. Without LRI,
 the Andreev current starts as $(V-V_0)^{3/2}$, whereas with LRI, it starts 
to increase faster than any power law. This can be viewed as a signature of LRI 
and could be tested in experiments.

In the conclusion, we discuss two possible applications, one for
 realistic fabricated JJ chains 
and the other for one-dimensional superconducting wires.
 For JJ chains, the interaction between island charges is screened
 but the methods used in this paper can be adapted to make at least qualitative
 predictions. For 1d wires, owing to the strong dependence of a phase slip 
 fugacity with temperature and wire width, and also for other reasons detailed below,
 it is not obvious that LRI can be detected  experimentally. Maybe they could
 for very long wires.


\section{Generalized Villain model and effective action with long range, $XY$ model
 in two dimensions}

\subsection{Imaginary time action}

We would like to study the influence of poorly screened Coulomb interactions 
in 1 dimension. One possibility, which has been widely used
 some time ago, mainly in dimensions two and three, would be to take the
 Hamiltonian for a BCS superconductor \cite{dePalo,Randeria}
 and to introduce Hubbard-Stratonovich transformations in the particle-hole
  and particle-particle channels in order to make the usual
 gap $\Delta = \vert \Delta \vert e^{i \varphi}$ appear. 
The amplitude fluctuations are integrated out, so are the phase of
 the density fluctuations, leaving us with a phase action only. 
This procedure has been used in Ref. \onlinecite{LobosGiamarchi}, 
except that they stayed at the RPA level and did not consider
 topological excitations for the phase $\varphi$ of the order parameter. 
Another way would be to start with a one-dimensional array
 of Josephson junctions, modelled by an ideal Josephson junction (JJ) 
and a capacity $C_J$ in parallel.
 In addition, on each island, the capacity to the ground has $C_g$ to be taken into account.
See Fig. 1. 

\begin{figure}[h] 
\epsfxsize 13. cm  
\centerline{\epsffile{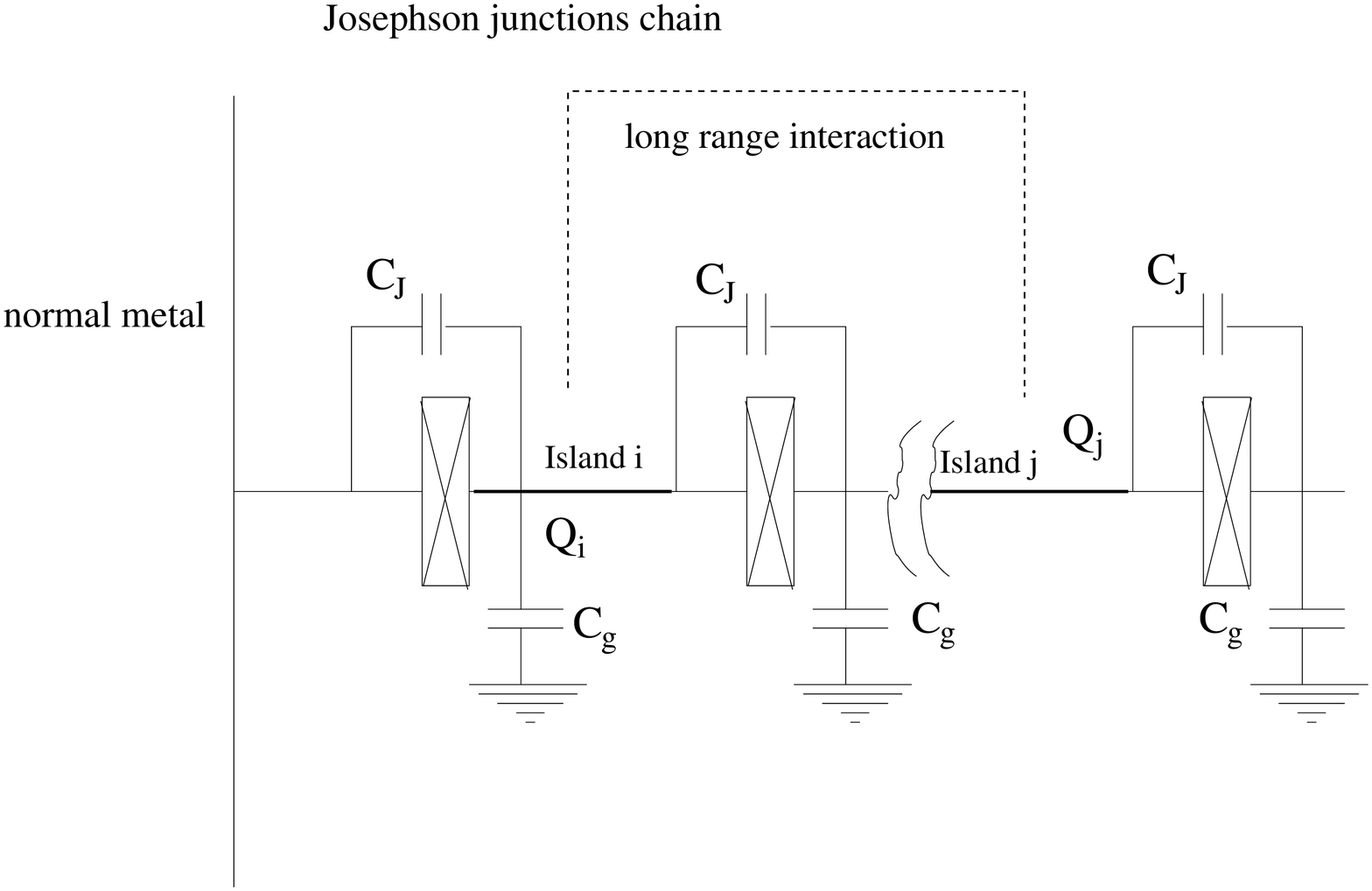}}
\caption{One-dimensional array of Josephson junctions
 with long-range Coulomb interactions.}
\end{figure} 
The Hamiltonian is
\begin{eqnarray}
H_{JJA} \, = \, \sum_{i} J \, \cos(\theta_{i+1}- \theta_i) + 
{1 \over 2} \sum_{i,j} n_i U_{i,j} n_j +
{A \over 2 C} \sum_{i,j} {q_i q_j \over \vert i-j \vert},
\label{HJJA0}
\end{eqnarray}
with $q_i = 2 e n_i$ and the interaction potential
 $U_{i,j} = 2 e^2 C_{i,j}^{-1}$
 and $C_{i,j}^{-1}$ is the capacitance matrix. The last 
term of Eq. (1) represents LRI, $A$ is a constant and $C$ a capacity. 
Usually, only the capacity of one junction $C_J$ and the capacity from an island to
 the ground $C_g$ are taken into account, the capacity matrix $C_{i,j}$ 
has the form $C_{i,j} = C_g + 2 C_J$ if $i=j$, 
$C_{i,j} = -C_J$ if $i= j \pm 1$, and $C_{i,j}=0$
 otherwise. It is reasonable to approximate the interaction 
potential\cite{GrabertDevoretproceedings} by the form 
\begin{eqnarray}
U_{i,j} = 2 {e^2 \over C_J} \lambda \, \exp-\Bigl({\vert i - j \vert \over \lambda}\Bigr),
\label{Uij}
\end{eqnarray} 
with $\lambda = \sqrt{C_J/C_g}$. The screening length is of the order $a \lambda$,
 with $a$ the interjunction spacing. Typically, for one-dimensional fabricated chains\cite{Chow} 
 $C_J \gg C_g$ but still $\lambda$ is of order $10$ to $30$, not more. 
LRI thus do not appear to be completely realistic for fabricated JJ chains,
 but for thin superconducting wires, the Coulomb interaction is badly 
screened\cite{Giamarchireview,Zaikinreview} and
 can be longer range. We shall thus take a Hamiltonian of the form
\begin{eqnarray}
H_{JJA} \, = \, \sum_{i} J \, \cos(\theta_{i+1}- \theta_i) + 
{1 \over 2} \sum_{i,j}
{q_i^2 \over 2 C} +
{A \over 2 C} \sum_{i,j} {q_i q_j \over \vert i-j \vert},
\label{HJJA}
\end{eqnarray}
 with $C^{-1}$ typically an effective capacitance 
to the ground. 
We can also see this Hamiltonian as a 1d analog of the following Hamiltonian
 for superconductors with fluctuations of the phase of the order parameter 
\begin{eqnarray}
H_S \, = \,
 \sum_{i} {\hbar^2 N_S^0 \over 2 m^*} \cos(\theta_{i+1} - \theta_i) 
 + \sum_i {1 \over 2} {m^* v^2 \over N_S^0} {\overline \rho}_i^2 \, 
+ \,\sum_{i,j} {\pi e^2 \over \epsilon}
 {{\overline \rho}_i {\overline \rho}_j \over \vert r_i - r_j\vert}.
\label{Hsupra1d}
\end{eqnarray}
In this case, $N_S^0$  is the superfluid density in 3d. 
${\overline \rho}$ is the charge density, 
$\theta$ will denote the phase of the order parameter. Its conjugate variable is
 $\Pi$, which is proportional to the charge $q$. 
When phase slips are neglected, the procedure to obtain an effective action for
 $\theta$ is standard and we just recall it here for completeness.

Writing $q_i = 2e \,n_i$ with $n_i$ integers, $\theta_i$ and $n_i$ are conjugate 
variables, $\lbrack n_i, \theta_i \rbrack \, = \, \delta_{i,j}$,
\begin{eqnarray}
H \, =\,
 \sum_{i,j} -J \cos(\theta_i - \theta_j) +
4E_0 n_i^2
 + 4 A E_0 \sum_{i,j}
{ n_i n_j  \over \vert i-j \vert},
\end{eqnarray}
with
$E_0={e^2 \over 2 C}$ the charging energy for nearest neighbor. 
Following Ref. \onlinecite{BradleyDoniach}, $T$ denoting
 the temperature and $\beta=1/(k_B T)$, the partition function is calculated 
in imaginary time, by discretizing in time steps of size 
$\hbar/\sqrt{2E_0 J}$. The system without LRI would be equivalent to a usual XY model
 with isotropic coupling. In the following,
 $\theta_{i,j}$ will no longer be operators but numbers; index
 $i$ from 1 to $N \equiv L/a$ denotes the position in real space and the second index
 $j$ from $1$ to $N^{\prime} \equiv 
\beta \sqrt{2E_0 J}$ the position on the imaginary
 time axis. Making $\hbar=1$, the
 partition function reads
\begin{eqnarray}
Z \, =  \, \sum_{\lbrace \theta \rbrace} \exp\biggl( - \sum_{i,j} 
\sqrt{J \over 2E_0} 
\lbrack \cos(\theta_{i+1,j} - \theta_{i,j})  + 
\cos(\theta_{i,j+1} - \theta_{i,j}) \rbrack\biggr),
\end{eqnarray}
where the sum is on all configurations $\lbrace \theta \rbrace$.

Now, in the spin-wave approximation, the cosines are expanded and the
Lagrangean $L$ and action $S$  take the form
\begin{eqnarray}
L \, &=& \, - \sum_{i=1}^N {1 \over 2} J (\theta_{i+1,\tau} - \theta_{i,\tau})^2 
 + 4E_0 n_i(\tau)^2 + 4AE_0 \sum_i \sum_{i^{\prime}} {n_i n_j \over \vert i-j \vert} 
 + \hbar \sum_i n_i(\tau) \partial_{\tau}  \theta_i(\tau),\\
S \, &=& \, -i \int_0^{\beta \hbar} L \, d \tau.
\end{eqnarray}

Applying a Fourier transform and integrating out the $\Pi_q = n_q$ field, 
\begin{eqnarray}
S_{eff} \, = \, \int_0^{\beta \hbar} 
\biggl\lbrack
\sum_i  \Bigl({1 \over 2}\Bigr)\,J
  \Bigl(\theta_{i+1}(\tau) - \theta_{i}(\tau)\Bigr)^2 
+  \int_q {1 \over 4E_0 + 4 A E_0 I(q)} \, 
(\partial_{\tau} \theta_q ) (\partial_{\tau} \theta_{-q}) 
\biggr\rbrack \,\,
d \tau,
\end{eqnarray}
where $I(q)$ is the Fourier transform of the Coulomb potential, 
regularized\cite{Giamarchireview} at very small
 $q$. $\partial_{\tau} \theta_q$ is in fact ${\hat \theta}_{q,j+1}
 - {\hat \theta}_{q,j}$, and ${\hat \theta}_{q,j}$ is the Fourier transform in space
 of $\theta_{i,j}$.  

Rescaling the imaginary time and counting it in units of $\hbar/\sqrt{2E_0 J}$,
 and setting $\hbar=1$ 
gives the effective action on a discrete space-time lattice
\begin{eqnarray}
S_{eff} \, = \, 
\sum_{i=1}^N \sum_{j=1}^{N^{\prime}} 
\Bigl({1 \over 2} \Bigr) \, \sqrt{{J \over 2E_0}}
 \Bigl(\theta_{i+1,j} - \theta_{i,j}\Bigr)^2 
+
 \sqrt{{J \over 8E_0}}
 \sum_q  {1 \over 1 +  A I(q)} \, \sum_{j=1}^{N^{\prime}} 
\bigl({\hat \theta}_{q,j+1} - {\hat \theta}_{q,j} \bigr)
\bigl({\hat \theta}_{-q,j+1} - {\hat \theta}_{-q,j} \bigr),
\end{eqnarray}
with $N^{\prime} \, = \, \beta \sqrt{2 E_0 J}$ and
 ${\hat \theta}$ the Fourier transform in space of $\theta_{i,j}$ defined as
\begin{eqnarray}
{\hat \theta}_{q,j} \, = \, {1 \over \sqrt{N}} \,
\sum_{n=1}^N \, e^{2 \pi i n } \theta_{n,j},
\end{eqnarray} 
for $q$ multiples of $2 \pi/N$.

Now, taking into account the periodic character of $\theta$, 
we can take the effective action
\begin{eqnarray}
S_{eff} \, =\, 
\sum_{i=1}^N \sum_{j=1}^{N^{\prime}} 
\sqrt{{J \over 2E_0}} \cos\bigl(\theta_{i+1,j} - \theta_{i,j}\bigr) 
+
 \sqrt{{J \over 8E_0}}
 \sum_q  {1 \over 1 + A I(q)} \, \sum_{j=1}^{N^{\prime}} 
\bigl({\hat \theta}_{q,j+1} - {\hat \theta}_{q,j} \bigr)
\bigl({\hat \theta}_{-q,j+1} - {\hat \theta}_{-q,j} \bigr).
\label{effectiveaction}
\end{eqnarray}

Having derived an effective action, 
we first turn to the case where the periodic character of the phase
 can be neglected. The approximation is good for large $J/E_0$.
 Cooper pairs feel very little Coulomb-blockade.
 The fugacity of phase slips goes
 essentially\cite{Zaikin} as $\exp(- 4 \sqrt{2J/E_0})$. 
Phase slips are neglected and a phase correlator
 derived. Owing to LRI, the system is no more a Luttinger liquid. Correlations 
ressemble some correlators appearing in one-dimensional quantum wires with low 
electron densities\cite{Schulz,Giamarchireview},
 sometimes dubbed ``Wigner crystals''.

\subsection{Spin waves}
In this section, we allow only smooth variations of $\theta$. 
It is then possible to expand the cosine term. This  gives immediately a spin-wave Hamiltonian. 
We check here the known spin wave dispersion relations.
\begin{eqnarray}
S_{eff} \,=\, \sum_{\omega} \sum_q 
\sqrt{J \over 8 E_0}
\, \Bigl\lbrack  q^2 +  {1\over 1 + A I(q)} \omega^2 \,\Bigr\rbrack 
\theta_{q,\omega} \, \theta_{-q,- \omega},
\end{eqnarray}
with $I(q) \simeq \ln(q^{-1})$ for small $q$. 
Setting $K= {1 \over 2 \pi} \sqrt{8 E_0 \over J}$, this results in 
\begin{eqnarray}
\langle \theta_{q,\omega} \theta_{-q,- \omega} \rangle \, =\, 
\,
{\pi K \over q^2 + {1 \over 1 +  A I(q)} \omega^2},
\label{thetaqw}
\end{eqnarray}
and thus
\begin{eqnarray}
G(x,\tau) \, = \, \langle \bigl(\theta(x,\tau)- \theta(0,0)\bigr)^2 \rangle \, =\, 
2 \pi K \,\, {1 \over \beta} \sum_{\omega_n} 
\int_{- \infty}^{\infty} 
 {1 - \cos(\omega_n \tau + q x) \over 
{q^2 + {\omega_n^2\over 1 + A I(q)}}} 
 \, {dq  \over \sqrt{ 2 \pi}}, 
\end{eqnarray}
where $\beta=1/(k_B T)$ and $\omega_n$ are the bosonic Matsubara frequencies. 
For distances $x$ and times $\tau$ not too long, so that the long range term 
 does not influence the behavior, the 
correlations are those of a Luttinger liquid
 $G(x, \tau) \, =\, K \, \ln\Big({\sqrt{x^2 + \tau^2} \over \alpha}\Bigr)$,
 where $\alpha$ is a short range cutoff (corresponding to finite  bandwidth),
 whereas for very large times\cite{Giamarchireview},
\begin{eqnarray}
G(x,\tau) \, = \, K \sqrt{2 A} \,
\ln^{3/2}\biggl( 
{\sqrt{x^2+ \bigl\lbrack\sqrt{2A} \, \tau \sqrt{\ln\tau}\bigr\rbrack^2} \over \alpha}
\biggr),
\end{eqnarray}
if $\tau$ is counted in units of $\hbar/\sqrt{2 E_0J}$.
The correlation function of $e^{i \theta}$ for spin-waves read 
\begin{eqnarray}
g(\tau) \, \equiv \, \langle e^{i \theta(0,0)} e^{- i \theta(0, \tau)}
 \rangle \, = \,  
 \exp\Biggl(
 - B \biggl\lbrack \ln(\tau) + {1 \over 2} 
\ln\Bigl(\ln \, \tau\Bigr) \biggr\rbrack^{3/2}\Biggr),
\end{eqnarray}
with $B= K\sqrt{{A \over 2}}$.
We would like to calculate the full correlation function, not only in
 the spin wave approximation, but including phase-slips. 
This turns out to be more intricate, we will neglect all factors
 that vary slower than any power law, such as, for example,
 logarithmic corrections or terms as $\exp(- C_0 \sqrt{\ln \,\tau})$ 
with $C_0$ a constant. 
Within this approximation, we have the spin-wave correlator
\begin{eqnarray}
g(\tau) \simeq \,
 \exp\Biggl( - B \biggl\lbrack \ln(\tau) \biggr\rbrack^{3/2}\Biggr).
\end{eqnarray}
\subsection{Phase slip Hamiltonian}

Now, we take into account phase slips, that is vortices for the 
 $XY$ model in 1 space  and 1 time dimension.  
The procedure to split the action into two parts, one due to spin-waves 
and the other to vortices is standard without LRI, however we recall it here with 
some details to point out the influence of LRI.

We use the model of Villain 
 and transform the exponential of the cosine term in (\ref{effectiveaction}) 
as the sum over $n$ integers
 but with a quadratic action\cite{Villain}, involving $\theta$ and $n$. 
The space direction will be denoted as $x$ and the imaginary time direction as $y$.
The long range interaction occurs only in the $x$ direction, for bonds
 directed in the $y$ direction. 
Let us denote by 
\begin{eqnarray}
F(q) \, =\, {1 \over 1+ A I(q)}.
\end{eqnarray}
As a consequence, the Coulomb term, i.e. the second term
 on the r.h.s of Eq. (\ref{effectiveaction}), has to be modified from
$\sum_q  F(q) \sum_{j=1}^{N^{\prime}}
\bigl({\hat \theta}_{q,j+1}- {\hat \theta}_{q,j}\bigr)
\bigl({\hat \theta}_{-q,j+1}- {\hat \theta}_{-q,j}\bigr)$
to
$\sum_{q} F(q) \, \sum_j 
\Bigl\lbrack 
\sum_l  \bigl( \theta_{l,j+1} - \theta_{l,j} - n^y_{l,j} \bigr) e^{i q l} 
\Bigr\rbrack
\Bigl\lbrack 
\sum_r  \bigl( \theta_{r,j+1} - \theta_{r,j} - n^y_{r,j}\bigr) e^{i q r} 
\Bigr\rbrack,$
where $n_{r,j}^{\alpha}$ denotes the ($2 \pi$ times) integer 
attached to the bond which goes from site $(r,j)$ to site $(r,j+1)$, 
and pertaining to the $y$ direction. The $n_{q,k_y}^y$ are simply
 the Fourier transforms of the $n_{r,m}^y$. The 
lattice spacing will be denoted $a$, and 
${\bf e}_x$ (${\bf e}_y$) will be the unit vector in the $x$ ($y$) direction. 

The part not involving the $n_k$'s is of the type
$\sum_{q} F(q) \sum_{k_y} \theta_{q.k_y} \, \theta_{-q,-k_y}$
 so that finally the total action is of the form
\begin{eqnarray}
\sum_{\alpha=x,y} \sum_{{\bf k}} \Bigl(F(q) \delta_{\alpha,y} + 
\delta_{\alpha,x} \Bigr)
 \, \vert i K_{\alpha} \theta_{ \bf k} - 2 \pi n_{\bf k}^{\alpha}\vert^2 + 
 \epsilon \vert\theta_k \vert^2,
\end{eqnarray}
where
\begin{eqnarray}
K_{\alpha} \,=\, 2\,sin({\bf k}.{\bf a}^{\alpha}/2),
\end{eqnarray}
and ${\bf a}^x= a {\bf e}_x$, ${\bf a}^y= a {\bf e}_y$ where
${\bf e}_x$ and ${\bf e}_y$ are the unit lattice vectors. 
$\epsilon$ is an infinitesimally small positive quantity. 
From now on, we will denote $q=k_x$ so that ${\bf k}$ has components $(k_x,k_y)$. 
The only difference with the normal XY model is that the coupling constant
 in the $y$ direction has a dependence in $k_x$ for small $k_x$ (long wavelengths). 

Setting ${\cal A}_x = \sqrt{{J \over 2 E_0}}$ and ${\cal A}_y = {\cal A}_x F(q)$,
\begin{eqnarray}
S_{eff} \, =\, \sum_{{\bf k}}
 \sum_{\alpha=x,y} \, {\cal A}_{\alpha} \,
 \vert  K_{\alpha} \theta_{{\bf k}} - 2 \pi n_{{\bf k}}^{\alpha} \vert^2.
\end{eqnarray}
Diagonalizing this action yields the decomposition into a continuous field 
${\tilde \theta}_k$ and a
 discrete field $n_{{\bf k}}$.
 The action must be of the form
\begin{eqnarray}
S_{eff} \, &=& \, S_{\theta} + S_V, \\
S_{\theta} &=& \sum_k  C(k) {\tilde \theta}_k^* \tilde{\theta}_k,\\
S_V &=& \sum_{\alpha,\gamma,k} n_{k,\alpha} n_{k,\gamma}^*D_{\alpha,\gamma}(k), 
\end{eqnarray}
where $\alpha$ and $\gamma$ take 
 two  values, $x$ and $y$. The quantities $D_{\alpha,\gamma}(k)$ 
and $C(k)$ are functions of 
${\bf k}$. 
The linear transformation is given explicitly by
\begin{eqnarray}
{\tilde \theta}_k \, &=& \, \theta_k + 2 i \pi 
{\sum_{\alpha}{\cal A}_{\alpha} K_{\alpha} n_{k,\alpha}  \over 
\epsilon + \sum_{\alpha}{\cal A}_{\alpha} K_{\alpha}^2},\\
C(k) \, &=& \, \sum_{\alpha = x,y} 
\Bigl( F(k_x)\delta_{\alpha,y} + \delta_{\alpha,x}\Bigr),\\
D_{\alpha,\gamma}(k) \, &=& 
\, 4 \pi^2 {{\cal A}_x {\cal A}_y   \over {\cal A}_x K_x^2 + {\cal A}_y K_y^2} \,
 K_{\alpha} K_{\gamma}
 \, (2\delta_{\alpha,\gamma}-1),
\end{eqnarray}
 and $\delta_{\alpha,\gamma}=1,$ if $\alpha=\gamma$ and zero otherwise.

Next, $S_V$ is rewritten in terms of vortices defined as
\begin{eqnarray}
q_k \, = \, i(K_y n_{k,x} - K_x n_{k,y}),
\end{eqnarray}
which yields the vortex Hamiltonian
\begin{eqnarray}
H_V \ =\, 4 \pi^2\sum_{k \not= 0} {\cal A}_x {\cal A}_y {q_k q_{-k}  \over 
{\cal A}_x K_x^2 + {\cal A}_y K_y^2}.
\end{eqnarray}
The apparent divergence at $k=0$ is in fact absent because of
 charge neutrality $(q_{k=0} = \sum_i q_i \, =\,0)$. 
Going back to real space, denoting $N_{tot}=N \, N^{\prime}$, the total number of sites:
\begin{eqnarray}
q_{\rho} \, = \, {1 \over \sqrt{N_{tot}}} 
\sum_{{\bf k}} q_{{\bf k}} e^{i {\bf k}.{\bf R}_{\rho}},
\end{eqnarray} 
enables to rewrite the vortex Hamiltonian as 
\begin{eqnarray}
H_V \ =\, \sum_{\rho,\rho^{\prime}} V_{\rho,\rho^{\prime}}
 q_{\rho} q_{\rho^{\prime}},
\end{eqnarray}
with, taking into account values of ${\cal A}_x$, ${\cal A}_y$ and $F(q)$,
\begin{eqnarray}
V_{\rho,\rho^{\prime}} \, =\, 
- 4 \pi^2 \sqrt{{J \over 2 E_0}} \, 
{1  \over N_{tot}} \sum_{{\bf k}} 
\Bigl\lbrack 1 - \cos\bigl({\bf k} .
 ({\bf R}_{\rho} - {\bf R}_{\rho^{\prime}}) \bigr)  
\Bigr\rbrack \, 
\Bigl\lbrack { 1 \over \lbrack 1 + A I(k_x) \rbrack  K_x^2 + K_y^2}
\Bigr\rbrack.
\end{eqnarray}
First, the self-interaction $V_{\rho,\rho}$ has to be withdrawn,
 but it is not so crucial because of charge neutrality 
($\sum_{\rho} q_{\rho}=0$).
Second, the vortex Hamiltonian as it stands is incomplete. 
The vortex fugacity term, of the form
 $\sum_{\rho} (\ln \,y) q_{\rho}^2$ is missing, with $\ln \,y$ being the fugacity
 of a vortex. In fact, one should start with a generalized Villain model
\cite{JKKN}, introduced 
by Jos\'e, Kadanoff, Kirkpatrick and Nelson, 
hereafter denoted by JKKN.
 As noted in Ref. \onlinecite{JKKN}, 
such a term is naturally generated by real space (Migdal-Kadanoff) renormalization. It
 is therefore necessary to reinstall this term.
 Next, we look at the shape of the interaction between vortices.

\subsection{Two regimes }

Although it is known that, with LRI, the vortex-vortex coupling 
for large separations \cite{MPAFisher} (much larger than the lattice constant)
 goes only as 
in $\sqrt{\ln \vert {\bf R}_{\rho} - {\bf R}_{\rho^{\prime}}\vert }$, we need to derive a
 quantitative expression.  
 The behavior is dominated
 by the behavior of the integral for
 small $\vert {\bf k} \vert$. For theses small wave-vectors,
 $K_x \simeq a k_x$ and  $K_y \simeq a k_y$, $a$ being the lattice
 constant. We can take ${\bf R}_{\rho}= {\bf 0}$ and
 ${\bf R}_{\rho^{\prime}} =
\rho \, \cos(\nu) {\bf e}_x + 
\rho \, \sin(\nu) {\bf e}_y $.
\begin{eqnarray}
V_{\rho,\rho^{\prime}} \, &\simeq&
 \, - 4 \pi^2 I, \nonumber \\
I \, &=& \, \sqrt{{J \over 2 E_0}} \,
\int \!\! \int
{ 1 - 
\cos\Bigl( k_x \rho^{\prime} \cos(\nu) + k_y \rho^{\prime} \sin(\nu) \Bigr)
\over
\Bigl\lbrack 
\Bigl(1 + A \, I(k_x) \Bigr) k_x^2 \, + \, k_y^2 
\Bigr\rbrack} \, d^2{\bf k}.
\label{vrhorhoprime}
\end{eqnarray} 
The cosine term will be close to $1$ only for $k_x$ much lower
 than $1/\lbrack \rho^{\prime} \cos(\nu)\rbrack$, and
  $k_y$ much lower
 than $1/\lbrack \rho^{\prime} \sin(\nu)\rbrack$.
 To extract the main behavior for large distances $\rho^{\prime}$,
 we can replace it by the integral
\begin{eqnarray}
I \, =\,  4 \, \sqrt{{J \over  2 E_0}}
\int_{(\rho^{\prime} \cos(\nu))^{-1}}^{k_m} \!\!
\int_{(\rho^{\prime} \sin(\nu))^{-1}}^{k_m} 
{1 \over
\Bigl\lbrack
\bigl\lbrace  1 + A \, \ln(k_x^{-1}) \bigr\rbrace
 \, k_x^2 \, + \, k_y^2 
\Bigr\rbrack}\, dk_x \, dk_y,
\end{eqnarray}
where $k_m$ is a wave-vector much larger than 
 ${1 \over \rho^{\prime}}$, but still sufficiently small 
that the expansion 
 of $K_x$ and $K_y$ for small $k$ holds.
The integral reads
\begin{eqnarray}
I \, &=&\, 
4 \sqrt{{J \over  2 E_0}} \int_{1/\bigl(\rho^{\prime} \cos(\nu)\bigr)}^{k_m}
{1 \over \sqrt{1+A \, \ln(k_x^{-1})} \, k_x} \, \nonumber \\
&\,& \Biggl\lbrack
{\rm atan}\biggl({k_m \over k_x \sqrt{1+ A \ln(k_x^{-1})}}\biggr) -
{\rm atan}\biggl({1\over \rho^{\prime} \sin(\nu)
 k_x \sqrt{1+ A \ln(k_x^{-1})}}\biggr) 
\Biggr\rbrack \,\,
dk_x.
\end{eqnarray}

As $\rho^{\prime}$ gets large, for
 $\nu$ different from $0$ and $\pi/2$,
 $\rho^{\prime} \cos(\nu) /\bigl\lbrack 
A \sqrt{\ln(\rho^{\prime})} \bigr\rbrack$ goes to infinity,
 so the first ${\rm atan}$ term in the bracket is close to $\pi/2$ whereas
 the second ${\rm atan}$ term remains much smaller than $1$, provided
 ${\rm tan}(\nu)$ is not too small. We can thus safely replace 
the term into brackets by $\pi/2$.
The integral on $k_x$ is then readily performed and is in
 $A^{-1} \sqrt{1 + A \, \ln(k_x^{-1})}$. 
The limit $A=0$ is singular. It is better to discuss limiting cases.

Case 1:
$\bullet$ 
 $\ln\Bigl(\rho^{\prime}  \cos(\nu) \Bigr) \ll A^{-1}$,

This corresponds to distances where the decrease of the correlations due to
 long range unscreened Coulomb interactions still has not taken place. 
We have then
\begin{eqnarray}
V_{\rho,\rho^{\prime}} \sim 2 \pi \,\sqrt{{J \over 2 E_0}}
 \ln\Bigl(\rho^{\prime} \cos(\nu)\Bigr).
\end{eqnarray}

Case 2:
$\bullet$ 
 $\ln\Bigl(\rho^{\prime}  \cos(\nu) \Bigr) \gg A^{-1}$,

This corresponds to distances where the decrease of the correlations due to
 long range unscreened Coulomb interactions has taken place.

We have then
\begin{eqnarray}
V_{\rho,\rho^{\prime}} \sim 2 \pi \,\sqrt{{J \over 2 E_0}}
\sqrt{ \ln\Bigl(\rho^{\prime} \cos(\nu)\Bigr)/A}.
\end{eqnarray}
Thus, at very large distances, there is no longer sufficient interaction between vortices
 to hold them tied together, whatever the short range term is \cite{MPAFisher}.
 Mathematically the BKT transition is expected to
 disappear. However, we need to put this on a more quantitative basis.

Furthermore, in real systems, the Coulomb interactions remain unscreened only
 up to a certain screening length
 $\xi_e$, which can be large but is not infinite. Also, the magnitude of the
 LRI Coulomb term may be much smaller than the one of the local Coulomb term. 
In order to determine the phase diagram and the phase correlations, we need to implement
 a real space renormalization group scheme analog to the one developped in 
Refs. \onlinecite{BerezinskiiKosterlitzThouless,JKKN}. Care has to be taken since LRI
 will eventually dominate under the RG group. The last term in Eq. (2) is more relevant
 than the second one. However, under the RG fow, it will take some time for the
 last term (LRI) to
 overcome the effect of the local one (i.e. the second term in Eq. (2)). The situation 
is reminiscent of LRI in one-dimensional quantum wires with very low electron densities; 
there are important differences however. There will be remnants of the BKT transition. 
To proceed, we perform a RG in real space closely following Kosterlitz'original paper
\cite{BerezinskiiKosterlitzThouless}.


\section{Renormalization group in real space}

A quick look at the situation is that, owing to the Coulomb potential,
 at very large distances $R$, the interaction between vortices goes
 like $\sqrt{\ln \rho}$ (instead of $\ln \rho$ in the usual case) , 
and is too weak to overcome the entropy term for placing a free vortex, which goes 
as $\ln \rho$. Nevertheless, the last terms in Eqs. (\ref{HJJA},\ref{Hsupra1d})
 can be small compared 
to the second ones, even if they are less relevant in RG sense, their initial value
 can be much smaller. Thus, it makes sense to treat them as a small perturbation,
 at least initially. This leads us to investigate how the usual KT RG flow is modified.

\subsection{Modified KT equations}
Here, the partition function has a form which is slightly different from
 the usual KT flow. The gas is neutral, without electric field.
 $n$ will be the number of ``$+$'' charges.
\begin{eqnarray}
Z \,=\, 
\sum_n {1 \over n!^2} \, {\cal K}^{2n} \, \exp \Bigl\lbrace
\sum_{i,j} {\cal J} {\cal F}(r_i - r_j)  q_i q_j 
\ln \Bigl\vert {r_i - r_j \over \tau} \Bigr\vert \,\,\Bigr\rbrace,  
\end{eqnarray}
where ${\cal J}$ is a coupling constant, 
${\cal J} = 2 \pi \sqrt{{J \over 2 E_0}}$ and 
${\cal F}$ is a function that behaves as $1$ for not too large 
$r_i - r_j$ but decays as $\ln \vert r_i - r_j \vert^{-1/2}$ for very large 
$\vert r_i - r_j \vert$. A model ${\cal F}$ can be 
\begin{eqnarray}
{\cal F}(r_i-r_j) \, = \, \Bigl\lbrack 1 + \sqrt{A \, \ln \vert r_i - r_j \vert} 
\Bigr\rbrack^{-1}.
\label{fcaleq}
\end{eqnarray}
$\tau$ is the short time cutoff and ${\cal K}$ the fugacity.

In the case where the long range term is much smaller than the short
 range one (at the beginning of the renormalization procedure only), 
we can however follow the same steps as Kosterlitz. The details are shown in Appendix A.
We express the RG equation, using the variables
\begin{eqnarray}
X &=& {\cal J} -2, \\
Y &=& \bigl(4 \pi {\cal K} \tau^2\bigr)^2, \\
z &=& \ln \,\tau.
\end{eqnarray}
Now, with the long range interaction, this becomes
\begin{eqnarray}
{d X \over d z} \, &=&\,  - Y - {\cal B} \sqrt{z},
\label{firstflowequation}\\
{dY \over dz} \, &=&\, -2 XY,
\end{eqnarray}
with ${\cal B} = {4 \over 3} \sqrt{A}$. Note that ${\cal B}$ can be much smaller than one. 
Now, because ${\cal B}$ is  non-zero, the line $Y=0$ is
no longer a line of (stable or unstable) fixed points.

\subsection{Physical description of the flow}

The flow is no longer autonomous, as soon as ${\cal B} \not=0$. 
We expect a deformation of the usual flow. In particular, the line
$Y=0$ which was a line of fixed points in regular KT flow is no longer 
a true line of fixed points.

\begin{figure}[h] 
\epsfxsize 10. cm  
\centerline{\epsffile{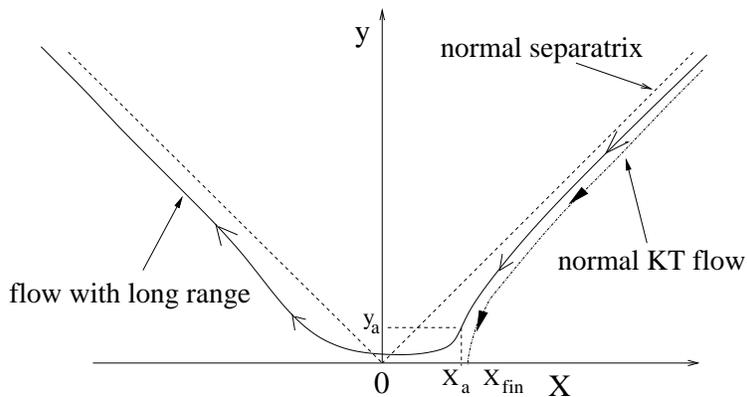}}
\caption{Modified KT flow in the presence of long-range interactions.
 Only a point which would correspond to the
 ``ordered'' phase of the usual KT diagram is shown. Here, because of
 long-range interactions, we always end in a completely disordered phase.}
\end{figure} 

If the LRI were absent, we could be either in the case where phase correlations
 are decaying exponentially (disordered phase) or in the ``ordered'' phase, where
 correlations decay as a power law. In the first case, adding the long range interactions
 will just make the phase correlation length smaller. The more interesting case is when
 we start from the ordered phase. We thus examine quantitatively
 how switching LRI drives us into the disordered phase and, in the next section,
 we shall examine what the phase correlations look like.

In the usual case, (without LRI), the flow is autonomous (i.e. does not depend explicitly on time) 
and trajectories are of the form
 $X^2 - Y \, = \, X_0^2 - Y_0 \,\equiv \, {\cal C}$,
 where $(X_0,Y_0)$ is the starting point. Setting $y \equiv \sqrt{Y}$,
 they are hyperbolae in the $(X,y)$ plane.
In the ``ordered'' phase, the 
flow on the invariant curves is:
\begin{eqnarray}
{dY \over Y \sqrt{Y+(X_0^2 - Y_0)}} \, = \, - 2 \, dz,
\end{eqnarray}
which integrates to 
\begin{eqnarray}
{2 \over \sqrt{{\cal C}}} \, \Biggl\lbrack
 \ln \biggl(\sqrt{{{\cal C} \over Y(z)}} + \sqrt{1 + {{\cal C} \over Y(z)}}\biggr) -
 \ln \biggl(\sqrt{{{\cal C} \over Y_0}} + \sqrt{1 + {{\cal C} \over Y_0}}\biggr)
\Biggr\rbrack \, = \, 2 (z-z_0).
\end{eqnarray}
$z_0$ is the initial ``time'' which can be set to zero.
Integration to obtain $X$ is also straightforward.

The important thing is that, as $z$ becomes very large, $Y$, 
which is related to
 the fugacity, tends to zero exponentially, while $X$ tends to a strictly positive
 value $X_{fin}$, see Fig. 2.

In our case, (with LRI), we can see what happens by performing the following approximation:

-  if ${\cal B} \sqrt{z}$ is much smaller than $Y$, we neglect it,

-  when, it becomes of the order of $Y$, we do the opposite and
 neglect $Y$ in the r. h. s. of
 the $dX/dz$ expression, Eq. (\ref{firstflowequation}). 

Thus, there are at least two stages. In the first stage, 
we have the usual KT flow whereas
 in the second stage the long range term comes into play. We shall see
 that there are three stages actually.
We now describe the first stage.
 Starting from a definite value of $Y$, by the regular KT flow, $Y$ will decrease. 
Suppose that $Y$ has decreased by a substantial amount,
 such that $Y$ will be much smaller than ${\cal C}$. This enables to replace
 $\sqrt{1 + {{\cal C} \over Y(z)}}$ 
by $\sqrt{{{\cal C} \over Y(z)}}$. Inverting the relation yields
\begin{eqnarray}
Y(z) \, = \, {4 {\cal C} \over D^2 \, \exp\Bigl(- \sqrt{{\cal C}} (z-z_0) \Bigr)},
\end{eqnarray}
with $D \, = \, \sqrt{{{\cal C} \over Y_0}} + \sqrt{1 + {{\cal C} \over Y_0}}$.
 ($ {{\cal C} \over Y_0}$ is not necessarily much larger than one).
The physics is that ${\cal C}$ is proportional to the initial distance to 
 the separatrix. It is proportional to $(T_c-T)/T_c$ in the thermal X-Y model. 
Now, we have to stop when $Y$ reaches the value ${\cal B} \sqrt{z-z_0}$, because we
 are no longer allowed to neglect the term ${\cal A} \sqrt{z}$ in
 Eq. (\ref{firstflowequation}).

This happens when
\begin{eqnarray}
{4 {\cal C} \over D}\, \exp\Bigl(- \sqrt{{\cal C}} (z-z_0) \Bigr) \, = \, {\cal B} 
\, \sqrt{z-z_0}.
\end{eqnarray}
To get the main feeling,
 we can show that the typical value of $z$ for which this happens is
 $z_1$, such that $z_1-z_0 \simeq {1 \over \sqrt{{\cal C}}}$. This 
corresponds to a typical imaginary time $\tau_1$, or corresponding lengthscale 
$l_1$ such that 
\begin{eqnarray}
\tau_1 \simeq l_1 \, =\, {1  \over {\cal B}^{1/\sqrt{{\cal C}}}}.
\end{eqnarray}
Neglecting $\ln(\ln)$ contributions, $Y_a$, the corresponding value of $Y$ is 
\begin{eqnarray}
Y_a \, =\, {\cal B} \sqrt{ {1 \over \sqrt{{\cal C}}} \,  \ln({1 \over {\cal B}})}.
\end{eqnarray}
$Y_a$ goes to zero as ${\cal B}$ goes to zero.
Now, we will enter the second stage.

We now describe this second stage, where
 the second term in Eq. (\ref{firstflowequation}) becomes larger than $Y$.
Then, for $z$ larger than $z_a$, we integrate the RG equations, 
neglecting the $Y$ term with respect to ${\cal B} \sqrt{z}$. 
\begin{eqnarray}
X \, = \, X_a \, - {2 {\cal B} \over 3} \, z^{3/2},
\end{eqnarray}  
and, using the other flow equation to obtain $Y$,
\begin{eqnarray}
\ln \, Y \, = \, \ln \, Y_a - 2 X_a z \, + {8 \over 15} {\cal B} \, z^{5/2}.
\end{eqnarray}

Then, $X$ will decrease and $Y$ will continue to decrease, till
$X$ passes the value $0$. $Y$ will be minimum
 for $X=0$ and then will start to increase. In fact, it increases faster
 than exponentially, so it will quickly become larger than
 $\bigl\lbrack {\cal B} \sqrt{z-z_0}\bigr\rbrack$,
 so that our approximation will fail. 
The value of $z_2$ for which this happens can be estimated, but anyway
 $z_2-z_0 \ll z_1-z_0$, corresponding to a length $l_2 = e^{z_2} \ll l_1$.  

Then, there is a third stage, where $Y$ becomes
 larger than  $\bigl\lbrack\sqrt{B} (z-z_0)\bigr\rbrack$.
Then, we can again neglect 
$\bigl\lbrack \sqrt{B} (z-z_0)\bigr\rbrack$ in front of $Y$.
We are then in the third regime which has analogies with the
 KT flow but on the disordered part of the transition.
This lasts till $Y$ has reached an appreciable value such that one is no longer in the vicinity
 of the critical point. Let $z_3$ be the corresponding value of $z$,
 and $l_3=e^{z_3-z_2}$. $l_3$
 will be also of order ${\cal B}^{-{1 \over \sqrt{{\cal C}}}}$. Since
 ${\cal C}$ is proportional to $T_c-T$, where $T_c$ is the transition temperature
 without LRI, it is tempting to conclude that,
 since ${\cal B}$ goes as $\sqrt{A}$, the correlation length with LRI 
$\xi^{\prime}$ behaves as 
\begin{eqnarray}
\xi^{\prime} =  D {\hbar \over \sqrt{ 2 E_0 J}}
 \exp\Bigl(\ln\bigl(A^{-1}\bigr) /2\sqrt{T_c-T}\Bigr),
\end{eqnarray}
with $D$ an unimportant constant of order one.
 It has the same form as the KT correlation length in the
 disordered phase (without LRI) but $\ln(A^{-1})$ 
replaces the usual\cite{BerezinskiiKosterlitzThouless} constant $b$. 
 Of course, $\xi^{\prime}$ tends to infinity as ${\cal B}$ tends to zero. 
The general shape of the ``flow diagram'' is best seen on Fig. 2.

Renormalization has to be stopped when $\tau$ becomes equal to $\beta \hbar$. 
Equating $\xi^{\prime}$ to $\hbar/(k_B T)$ yields a temperature $T^*$ given by
\begin{eqnarray}
T^* \, = \, (Dk_B)^{-1} \sqrt{2 E_o J} \, \exp\Biggl(-{\ln(A^{-1}) \over 2
 \sqrt{T_c-T^*}}\Biggr),
\end{eqnarray}
with $T_c$ the critical temperature in the absence of LRI.
If experiments are performed at a temperature larger than $T^*$, it will not
 be possible to see the influence of LRI.

Having the general shape of the ``flow-diagram'', 
we now turn to the determination of the phase correlation function.

\section{Correlation functions}
This section is a little technical. Just as the Hamiltonian decouples into a vortex part, the
 phase correlation function also do. Estimation of the spin-wave part is 
straightforward, but the estimation of the vortex part is more intricate. In section A,
 we first determine the contribution of one vortex to the phase correlator.
 This involves a vortex influence function $v(r)$. Comparison between the usual case and
 the case with LRI is emphasized. 

Then, in section B, average must be carried out over the possible configurations
 of vortices, using the Boltzmann factor with the vortex Hamiltonian. A Coulomb gas
 formulation is necessary. Distinction must be done between smooth variations of the 
vortex influence function $v(r)$,
 studied in section C, and abrupt variations (section D); 
there are thus two contributions. 
Summing both contributions gives the vortex part of the phase corrrelator. 
Adding the spin-wave part yields the total phase correlator 
in imaginary time. In section F, analytical continuation is
 performed to obtain it in real time. The phase correlator in real time is necessary
 for the calculation of the transport properties of JJ chains. 

For the determination of the correlation function, we use the method of JKKN. 
In the vortex Hamitonian, the core energy has to added. Such a term naturally arises from 
their Migdal-Kadanoff renormalization group. Therefore, we must use
 the generalized Villain lodel. The corresponding vortex Hamiltonian is
\begin{eqnarray}
H_V \, = \, \sum_{\rho,\rho^{\prime}} 
 V_{\rho,\rho^{\prime}} q_{\rho} \, q_{\rho^{\prime}} + 
(\ln \, y) \, \sum_{\rho}  q_{\rho}^2,
\end{eqnarray}
with $y$ the charge fugacity. 
The $q_{\rho}$ are integer charges living
 on the dual lattice of the original lattice. They represent vortices.
 The correlation function to be computed is 
$\langle e^{i \theta(x,\tau)} e^{- i \theta(0,0)}\rangle$. 
Written in Coulomb gas language, this
 correlation function disentangles into a spin-wave part and 
a vortex part. 
\subsection{Determination of the vortex influence function}

We closely follow the method \cite{Pelcovits,HeinekampPelcovits}
 of Pelcovits. 
The first thing to do is to determine what is the influence of a
 vortex sitting at ${\bf R}$ on the vortex part of the correlation function
 $g_V({\bf r} - {\bf r}^{\prime})$. 
The total correlation function is
\begin{eqnarray}
g_{tot}({\bf r} - {\bf r}^{\prime}) \, =\, 
\langle  e^{i \theta({\bf r})} e^{-i \theta({\bf r}^{\prime})} \rangle.
\end{eqnarray}
Following JKKN, $g_{tot}({\bf r} - {\bf r}^{\prime})$ 
is written as
 a product of nearest neighbor correlation functions, along a path ${\cal P}$
 that connects ${\bf r}$ to ${\bf r^{\prime}}$ on the real lattice. This is for the total
 correlation function (vortices plus spin-waves).   
After (Gaussian) integration 
 of the spin-wave degrees of freedom,
\begin{eqnarray}
g_{tot}({\bf r} - {\bf r}^{\prime}) \, = \,\
g_{SW}({\bf r} - {\bf r}^{\prime}) \, g_V({\bf r} - {\bf r}^{\prime}).
\end{eqnarray}
The vortex part, $g_V({\bf r} - {\bf r}^{\prime})$ 
is given by 
\begin{eqnarray}
g_V({\bf r} - {\bf r}^{\prime}) \, =\, 
\biggl\langle \exp\Bigl(i \sum_{{\bf R}} m({\bf R}) \, v({\bf R})\Bigr)
 \biggr\rangle,
\label{vortexpartofg}
\end{eqnarray}
with 
\begin{eqnarray}
v({\bf R}) \, = \, \sum_{{\bf R}^{\prime}} V({\bf R}- {\bf R}^{\prime}) 
\bigl\lbrack \eta^l({\bf R})- \eta^r({\bf R}^{\prime})\bigr\rbrack ,
\end{eqnarray}
where ${\bf R}$ are all the positions on the dual lattice, the function
 $\eta^l({\bf R})$ ($\eta^r({\bf R})$) 
is zero everywhere except on sites of the dual lattice that lie 
immediately on the left (right) 
of the path chosen joining ${\bf r}$ to ${\bf r}^{\prime}$. 
$V({\bf R}-{\bf R}^{\prime})$ is simply $V_{\rho,\rho^{\prime}}$, 
see Eq. (\ref{vrhorhoprime}). The 
brackets mean thermodynamical average  with respect to $H_V$.

This results directly from Gaussian integration, it can be understood
 as follows: 
the contribution of a vortex sitting at location 
 ${\bf R}$, and vorticity $m({\bf R})$ is 
$\exp\Bigl(i \, m({\bf R}) \, v({\bf R}) \Bigr)$.
The function $v({\bf R})$ depends on ${\bf r}$ and ${\bf r}^{\prime}$. 
It has also another
 physical interpretation. It is the phase generated at site ${\bf R}$ by
 a vortex-antivortex pair, where one vortex has $m=1$ and is sitting on ${\bf r}$
 and the antivortex ($m=-1$) is sitting on ${\bf r}^{\prime}$. 

The determination of $v({\bf R})$ has been done in the literature without the long
 range Coulomb interaction (LRI). This function will be called $u({\bf R})$,
 and the potential energy between vortices will be denoted as   
 $V^{0}_{\rho,\rho^{\prime}}$.

We explain how the LRI affect the results.
Following JKKN, without LRI,
\begin{eqnarray}
u({\bf R}) \, = \, 
\sum_{{\bf R}^{\prime}} G^{\prime}({\bf R} - {\bf R}^{\prime})\, 
 \Bigl(\eta^l({\bf R}^{\prime} - \eta^r({\bf R}^{\prime})\Bigr),
\label{udeR}
\end{eqnarray}
where $G^{\prime}({\bf R} - {\bf R}^{\prime}) = 
V^{0}_{{\bf R}_{\rho},{\bf R}^{\prime}_{\rho^{\prime}}} -
V^{0}_{{\bf R}_{\rho},{\bf R}_{\rho}} \,
 \simeq V^0_{\rho,\rho^{\prime}} - V^0_{\rho,\rho}$.
 See Fig. 3.
\begin{figure}[h] 
\epsfxsize 7. cm  
\centerline{\epsffile{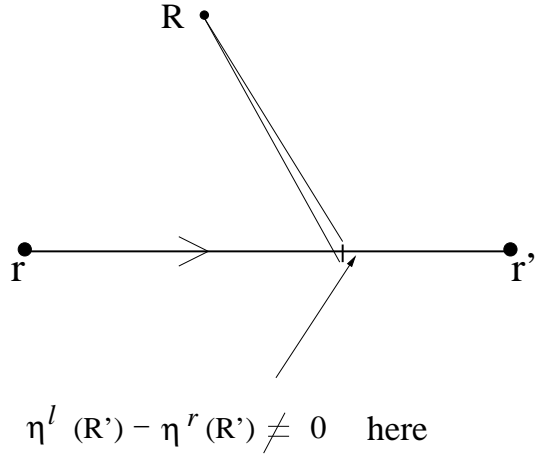}}
\caption{Determination of influence of vortex sitting at ${\bf R}$ on the correlation function
 $\langle e^{i \theta({\bf r})} e^{- i \theta({\bf r}^{\prime})}\rangle$.}
\end{figure} 

For not too large distances, the interaction between vortices is still logarithmic
 and we just recall the known results for completeness. For simplicity, we take a path
 joining ${\bf r}$ and ${\bf r}^{\prime}$ to be a straight line (see Fig. 3). 
Working in cartesian coordinates,
 and taking the origin at ${\bf r}$, the $x$ axis, along
 the line going from ${\bf r}$ to ${\bf r}^{\prime}$, 
the coordinates of ${\bf R}$, ${\bf r}$ and ${\bf r^{\prime}}$ will be
 $(x,y)$, $(0,0)$ and $(r^{\prime},0)$ respectively. 
The current integration variable
 ${\bf R}^{\prime}$ on the dual lattice immediately above the cut will be of coordinate
 $(x^{\prime},y^{\prime})$. 

\begin{eqnarray}
u({\bf R}) \, = \, lim_{y^{\prime} \rightarrow 0} 
\int_0^{r^{\prime}} {d \over dy^{\prime}} \,
 \ln \Bigl(\sqrt{(x-x^{\prime})^2 + (y-y^{\prime})^2}\Bigr) \, dx^{\prime} \, = \,
{\rm atan}(y/x) - {\rm atan}\bigl(y/(x-r^{\prime})\bigr), \nonumber \\
\end{eqnarray}
which gives 
\begin{eqnarray}
u({\bf R}) \, = \, \Phi({\bf r} - {\bf R}) - \Phi({\bf r}^{\prime} - {\bf R}),
\end{eqnarray}
with $\Phi({\bf Z} ) = {\rm atan}(y_z/x_z)$, if ${\bf Z} = (x_z,y_z)$.

Now, consider the LRI, at large distances
\begin{eqnarray}
V_{\rho,\rho^{\prime}} = 2 \pi \sqrt{{J \over 2 E_0}} 
\sqrt{\ln (\rho - \rho^{\prime})/A}.
\end{eqnarray}
The function $u({\bf R})$ at large distances must be replaced now by $v({\bf R})$. 
\begin{eqnarray}
v({\bf R}) \, = \, 
lim_{y^{\prime} \rightarrow 0} 
{2 \pi \over \sqrt{A}}\sqrt{{J \over 2 E_0}}\,
\int_0^{r^{\prime}} 
{d \over dy^{\prime}}
 \biggl( \sqrt{\ln \Bigl(\sqrt{(x-x^{\prime})^2 + (y - y^{\prime})^2}\Bigr)} \biggr)
\, dx^{\prime}. 
\end{eqnarray}
Integrating by parts, to extract the main contribution at large distances,
yields the result
\begin{eqnarray}
v({\bf R}) \, &=& \, 
{A^{-1/2} \over \sqrt{2}} \sqrt{{J \over 2 E_0}}\,
 \Biggl(-{1 \over \sqrt{\ln \Bigl((r^{\prime}- x)^2+ y^2\Bigr)}}
{\rm atan}({r^{\prime} - x \over y}) \, - \, 
{1 \over \sqrt{\ln \Bigl(x^2+ y^2\Bigr)}} {\rm atan}({x \over y})\Biggr) \,\,  \nonumber \\
&+& \,\,
{A^{-{1/2}} \over \sqrt{2}} \sqrt{{J \over 2 E_0}} \, F,
\end{eqnarray}
with $F$ a quantity which is smaller by a factor
 $\Bigl(\ln(x^2+y^2) \Bigr)^{-1}$ than the main one.
Thus, $v({\bf R})$ behaves\cite{MPAFisher}
 as $\Bigl\lbrack \ln( {\bf R}) \Bigr\rbrack^{-{1 \over 2}}$ for large
 ${\bf R}$. 

\subsection{High temperature expansion of the vortex part of the phase
 correlator}

From Eq. (\ref{vortexpartofg}),
 the average of $\exp\Bigl(i \sum_{{\bf R}} m({\bf R}) v({\bf R})\Bigr)$ has to be 
performed with the full vortex Hamiltonian (including core energy).
The Boltzmann weight of a particular configuration $m({\bf R})$ is given by 
$\exp\Bigl(A\bigl(m({\bf R})\bigr)\Bigr)$ with
 $A\bigl(m({\bf R})\bigr)\,=\,\sum_{{\bf R}} m^2({\bf R})\, \ln \, y \,
+ \sum_{{\bf R},{\bf R}^{\prime}} m({\bf R}) m({\bf R}^{\prime}) \, 
V_{{\bf R}-{\bf R}^{\prime}}$ and thus 
\begin{eqnarray}
g_V({\bf r} - {\bf r}^{\prime}) \, = \, 
{
\sum_{\lbrace m({\bf R})\rbrace}
\exp\Bigl(A\bigl(m({\bf R})\bigr) + i \sum_{{\bf R}}m({\bf R}) v({\bf R})\Bigr) 
\over 
\sum_{\lbrace m({\bf R})\rbrace} 
\exp\Bigl(A\bigl(m({\bf R})\bigr)\Bigr)
}. 
\end{eqnarray}
A high temperature expansion of $g_V({\bf r} - {\bf r}^{\prime})$ is needed. 
Applying the Poisson formula 
$f(m) = \sum_{p= - \infty}^{\infty} \int_{- \infty}^{\infty}
 f(\phi) \,e^{- 2 i \pi p \phi} \, d \phi$
 to the function
 $f(m) = \exp\Bigl(A\bigl(m({\bf R})\bigr)
 + i \sum_{{\bf R}} m({\bf R}) v({\bf R})\Bigr)$ and integrating
over the continuous auxiliary field $\phi$ enables to write it as
\begin{eqnarray}
g_V({\bf r} - {\bf r}^{\prime}) \, = \, 
{Z^{\prime} \over Z},
\end{eqnarray}
with
\begin{eqnarray}
Z^{\prime} \, =\, 
\sum_{\lbrace p({\bf R})\rbrace} 
\exp\biggl(-{ 1\over 2}
 \sum_{{\bf R},{\bf R}^{\prime}}
 \bigl\lbrack v({\bf R}) + 2 \pi p({\bf R}) \bigr\rbrack
 \bigl\lbrack v({\bf R}^{\prime}) + 2 \pi p({\bf R}^{\prime})\bigr\rbrack 
\langle m(0) m({\bf R}) \rangle_{cont}\biggr),
\end{eqnarray}
where the sum is over all configurations of $p({\bf R})$.
 $\langle m(0) m({\bf R})\rangle_{cont}$ is
 the correlation function calculated 
with weight $A\bigl(m({\bf R})\bigr)$ but as though the $m$ were 
continous variables. $Z$ is the same as $Z^{\prime}$ except that $v({\bf R})$ is absent.
However, because a switch from the $m({\bf R})$ variables to the $p({\bf R})$ variables 
(through the Poisson formula) has been made, now a low temperature expansion
 is needed. A high $T$ expansion in the $m({\bf R})$ variables transcribes into 
 a low $T$ expansion in the $p({\bf R})$ variables. In the absence of LRI, 
$\langle m(0) m({\bf R}) \rangle_{cont}$ 
has been calculated in Refs. \onlinecite{BerkerNelson,NelsonHalperin}. 
Even with the LRI, the correlation $\langle m(0) m({\bf R})\rangle_{cont}$ remains 
short range and can be approximated by nearest-neighbor only. 
Substituting the exact $\langle m(0) m({\bf R})\rangle_{cont}$ by a
 nearest neighbor interaction results in   
\begin{eqnarray}
Z^{\prime} \, = \, 
\sum_{\lbrace p({\bf R}) \rbrace} \exp\Bigl(-{ 1 \over 16 \pi^2} \sqrt{{2 E_0  \over J}} \,
\sum_{\langle{\bf R},{\bf R}^{\prime}\rangle}
\bigl(v({\bf R}) + 2 \pi p({\bf R}) - u({\bf R}^{\prime})
 - 2 \pi p({\bf R}^{\prime})\bigr)^2 \Bigr).
\end{eqnarray}
Following Refs. \onlinecite{Pelcovits,HeinekampPelcovits}, 
a low temperature expansion is performed, keeping only the essential
 diagrams. To simplify further,
 only the non-overhanging self-avoiding-walks (SAW's) joining
 ${\bf R}$ to ${\bf r}^{\prime}$ 
are kept which results in
\begin{eqnarray}
g_V({\bf r} - {\bf r}^{\prime}) \, = \,
\sum_n \,P_n({\bf r} - {\bf r}^{\prime})
e^{-n/4 \sqrt{{2 E_0 \over J}}} \,
 \exp\biggl({1 \over 8 \pi^2} \sqrt{{2 E_0 \over J}}\,
\int \!\int^{\prime} 
\vert \nabla_Rv({\bf R}) \vert^2 \, d^2{\bf R} \biggr),
\end{eqnarray}
where $\int\!\int^{\prime}$ means integration over all the plane except on the 
SAW. $P_n({\bf r} - {\bf r}^{\prime})$ is the number of non-overhanging
 SAW's of $n$ steps joining ${\bf r}$ and ${\bf r}^{\prime}$.

\subsection{Smooth variations of $v({\bf R})$}

The calculation of the correlation function
 splits into the integration on the smooth
 variations of the field $v({\bf R})$ and on the abrupt discontinuities caused by
 the different diagrams occuring in the high $T$ expansion of  $Z^{\prime}$. 
In our case, the situation looks a little more complicated because $v({\bf R})$ does not
 satisfy the Laplace equation. However, this causes
 no trouble because the volume contributions coming from $\Delta v$,
 where $\Delta$ is the Laplacian, will turn out to be very small. The main contributions come
 from the regions near the path ${\cal P}$ going from ${\bf r}$ to
 ${\bf r}^{\prime}$. 

More precisely, the weight
\begin{eqnarray}
{\cal W} \equiv \, \exp\biggl(-
{1 \over 8 \pi^2} \sqrt{{2 E_0 \over J}}\,
\int\!\!\!\int^{\prime} \, \Bigl(\nabla_{{\bf R}} v({\bf R})\Bigr)^2
 \, d^2{\bf R} \biggl),
\end{eqnarray}
 is to be evaluated, where the prime means integrations everywhere in the
 plane except on the diagrams. 

In the usual case, integration is made by parts
\begin{eqnarray}
\int \! \int^{\prime} \Bigl(\nabla_{{\bf R}} u({\bf R})\Bigr)^2  \, d^2{\bf R} \, &=& \, 
T_1 + T_2 + T_3, \\
T_1 &\equiv& - \int\!\!\int^{\prime}_{{\cal S}} u \, \Delta u \,  \, d^2{\bf R}, \\
T_2 &\equiv& \int_{\Gamma_{out}} u \, {\bf \nabla} u \, . d {\bf \sigma}, \\  
T_3 &\equiv& \int_{\Gamma_{in}} u \, {\bf \nabla} u \, . d {\bf \sigma}.  
\end{eqnarray}
${\cal S}$ denotes the shaded aera in Fig. 4, $\Gamma_{in}$ is
 a contour which encloses the cut (from ${\bf r}$ to ${\bf r}^{\prime}$) 
 at a small distance $b_0$ and 
$\Gamma_{out}$ is a circle of radius ${\cal R}$ where ${\cal R}$ will
 eventually be allowed to go to $\infty$. $d \sigma$ is the unit normal to the graph.
 On Fig. 4, it is simply the normal to the line joining ${\bf r}$ to ${\bf r}^{\prime}$. 
The first term $T_1$ vanishes since $\Delta u \,= \, 0$. 
The second term $T_2$ will tend to zero as ${\cal R}$ goes to infinity. 
The last term $T_3$ has to be calculated using Eq. (\ref{udeR}). Care has 
to be taken that
 $G^{\prime}({\bf R} - {\bf R}^{\prime})$ behaves only as $\ln({\bf R}-{\bf R}^{\prime})$
 for sufficiently large distances and this form of $G^{\prime}$ should not
 be used for small distances.

\begin{figure}[h] 
\epsfxsize 7. cm  
\centerline{\epsffile{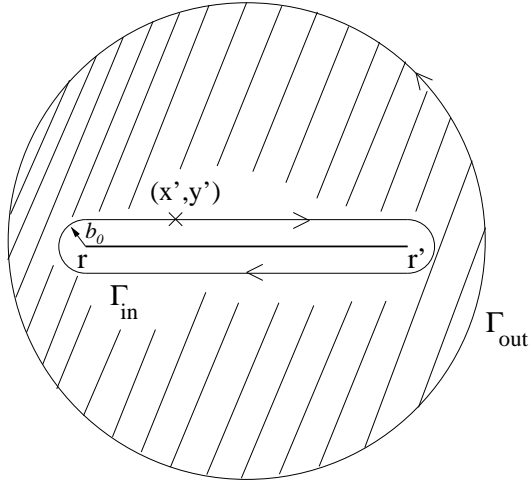}}
\caption{Determination of influence of the smooth variations of the field
 $u({\bf R})$ on the vortex part of the correlation function
 $\langle e^{i \theta({\bf r})} e^{- i \theta({\bf r}^{\prime})}\rangle$.}
\end{figure} 

With this in mind, 
taking ${\bf r} = {\bf 0}$ and thus
 $r^{\prime} = \vert {\bf r}^{\prime} - {\bf r} \vert$,  
\begin{eqnarray}
T_3 = \, 2 \, \int_0^{r^{\prime}} \, 
\Bigl\lbrack {\rm atan}({y^{\prime} \over x^{\prime}}) - 
{\rm atan}({y^{\prime} \over r^{\prime} - x^{\prime}})
\Bigr\rbrack\, 
 {d \over dy^{\prime}} 
\Bigl\lbrack {\rm atan}({y^{\prime} \over x^{\prime}}) - 
{\rm atan}({y^{\prime} \over r^{\prime} - x^{\prime}})
\Bigr\rbrack \, \, dx^{\prime}.
\label{T3}
\end{eqnarray}

The factor $2$ comes from the fact that integration above and below the cut
 ${\bf r},{\bf r^{\prime}}$ has to be performed (see Fig. 4).
We take the integration path a finite amount $b_0$ above (or below) the cut to avoid 
spurious divergencies (see above). Then $b_0$ will be unimportant. 
The angle difference in the difference of the two ${\rm atan}$ 
functions Eq. (\ref{T3}) 
will correspond
 to the change in angle, taken to be $\pi$. 
Dropping the spurious $\ln (b_0^2)$ terms yields\cite{deGennes} exactly
 $T_3 \, = \, 4 \pi \ln\Bigl(\vert {\bf r}^{\prime} - {\bf r} \vert \Bigr)$. 

Now, we turn to the case of LRI. The analog of the terms $T_i$, $i=1$ to $3$ will
 be denoted as $T^{\prime}_i$ and $u$ has to be replaced by $v$. 
Calculations are tedious and relegated to the
 appendices. It turns out that $T^{\prime}_1$ and $T^{\prime}_2$
 are completely negligible whereas $T^{\prime}_3 = 
 B_1 \, \ln\Bigl\vert \ln \vert {\bf  r} - {\bf r}^{\prime} \vert  \Bigr\vert$,
 with $B_1$ a constant. 
Even after taking the exponential of 
 this will only lead to logarithmic factors for the correlation functions and
we thus omit this factor.

\subsection{Abrupt variations of $v({\bf R})$, graph contributions}

Here, the part due to abrupt variations of the field $v({\bf R})$, 
 is the same as without LRI sum on SAW joining
${\bf r}$ and ${\bf r^{\prime}}$ and the result is analogous to the usual case, namely,
 the contribution, called $g_{V \, SAW}$ reads:
\begin{eqnarray}
g_{V \, SAW} \, \simeq \, {
\exp\Bigl(- {\vert{\bf r} - {\bf r}^{\prime}\vert  \omega_p\over \xi}\Bigr) 
\over \sqrt{\vert{\bf r} - {\bf r}^{\prime}\vert \omega_p}},
\label{OZnormal}
\end{eqnarray}
$\omega_p$ is the plasma frequency $\sqrt{8 E_0\,J}/\hbar$. 

Being the result of a high-T expansion, this
 expression corresponds to short correlation length $\xi$. It is probably not valid
 for very large $\xi$. In the limit $A \rightarrow 0$, the usual divergence close
 to the BKT transition must be retrieved.   

\subsection{Final form}

The spin-wave part, the vortex part coming from
 smooth variations of $v({\bf R})$ and the vortex part coming from abrupt 
  variations of $v({\bf R})$ just need to be multiplied.
Thus we have, with $r-r^{\prime} = \vert {\bf r} - {\bf r}^{\prime}\vert$,
\begin{eqnarray}
g(r-r^{\prime}) =
 \exp\Bigl(-B \bigl\lbrack \ln \vert (r-r^{\prime}) \omega_p 
\bigr\rbrack \vert^{3/2} \Bigr) \, 
\exp(B_1 \,\ln\,\, \ln\vert (r-r^{\prime}) \omega_p \vert )\, 
{\exp(- {\vert (r -r^{\prime}) \omega_p \over \xi}) \over \sqrt{(r-r^{\prime}) \omega_p}},
\end{eqnarray}
with $B_1$ a constant. 
Contrary to the usual case, the part of the vortex contribution corresponding 
to smooth variations of $v(R)$ no longer cancels the spin-wave contribution.
 Essentially
\begin{eqnarray}
g(r-r^{\prime}) \, = \,
\exp\biggl(-{\vert r-r^{\prime}\vert \omega_p\over \xi}\biggr) \, 
\exp\biggl(-B \Bigl\vert \ln \vert (r -r^{\prime}) \omega_p
 \vert \,\Bigr\vert^{3/2}\biggr)
\, \vert (r-r^{\prime}) \omega_p \vert^{-1/2},
\end{eqnarray}
where we systematically dropped all contributions going slower than any
 power law. We see that, contrary to what happens in the usual case,
 the total correlator
 is not of the Ornstein-Zernike type. However, in the limit $B \rightarrow 0$,
 the Ornstein-Zernike form is retrieved. 
\subsection{Analytic continuation}

It is now necessary to obtain the phase correlator in real time, and  then
 in real frequency. We need
 to perform an  analytical continuation $\tau \longrightarrow \,\, i\, t$. 
We do not attempt to get the short time behavior. 
This would correspond to high voltages. 
At long times,
\begin{eqnarray}
p(t) \, \propto \, \exp(- \vert \omega_p t \vert \xi^{-1}) \, \vert \omega_p t \vert^{-1/2} \,
\exp\Bigl(-B \, \bigl\vert \ln \vert \omega_p t \vert \bigr\vert^{3/2}\Bigr),
\end{eqnarray}
where $\propto$ means ``proportional to''. 
Normalization of $p$ would require more information
 on the shape of $p(t)$ for shorter times. This yields
 in real frequencies $\Omega$, for small $\Omega$'s (with respect to $\omega_p$)
\begin{eqnarray}
P(\Omega) \, \propto \, \Theta(\Omega - \omega_p\xi^{-1}) 
\exp\Bigl(-B \,
 \bigl\vert \ln \vert \Omega/\omega_p - \xi^{-1} \vert \bigr\vert^{3/2}\Bigr)
 \, (\Omega/\omega_p -\xi^{-1})^{-1/2},
\end{eqnarray} 
where $\Theta$ is the Heaviside function.
 For larger $\Omega$, this form is not valid. One would need the short time behavior
 of $p(t)$, which we do not have explicitly. However, on those short timescales, the long
range part of the interaction will play a minor role. Therefore, in order
 to perform calculations, a reasonable 
assumption is that, as without LRI,
 $P(\Omega)$ is negligible\cite{Falci} for $\Omega$ larger than $\kappa \omega_p$,
 with $\kappa=2$. 
Enforcing the normalization $\int_0^{\infty} P(\Omega) \, d \Omega =1$,
 we thus have the approximate form
\begin{eqnarray}
P(\Omega) \, = \, p_0 \, \omega_p^{-1} \, 
\Theta(\Omega/\omega_p - \xi^{-1}) \,
 \exp\biggl(- B \bigl\vert \ln \vert \Omega/ \omega_p - \xi^{-1} \vert \bigl\vert^{3/2}
\biggr)  \, (\Omega / \omega_p - \xi^{-1})^{-1/2},
\end{eqnarray} 
with $p_0 =  \bigl\lbrack 2 {\cal L}(2^{3/2}B,\sqrt{\kappa-\xi^{-1}})\bigr\rbrack^{-1}$, and
 the function ${\cal L}$ is defined by 
${\cal L}(b,y) = \int_0^y 
\exp\Bigl(-b \Bigl\vert \ln\vert x \vert \Bigr\vert^{3/2}\Bigr) \, dx$.

\section{Physical Applications}

Having obtained the phase correlator in real time, we
 then use it for the calculation of the Andreev current.
 A normal metal is placed on the left of the Josephson junction chain and biased  
by a voltage $V$, below the gap $\vert \Delta \vert$, see Fig. 1.

\subsection{Andreev average current}

The average Andreev current flowing from the metal into the superconductor
 can be readily expressed as a function of the distribution of phase modes
 $P(\Omega)$. The case $P(\Omega) = \delta(\Omega)$ corresponds to a usual
 BCS superconductor, and BTK results\cite{BTK} are retrieved. The case where
 $P(\Omega)$ is given by the Luttinger liquid theory or usual
 2d $XY$ model has been studied in Refs. \onlinecite{Falci,Devillard}. 
The contact between the metal and the JJ chain
 can range from tunnel to perfect contact.
 Following the Keldysh formalism,
 and performing a nonperturbative expansion, to all orders in 
the tunneling amplitude, one can describe all situations\cite{MartinRodero}. 
Without phase fluctuations, the procedure has been
 shown to be exactly equivalent to solving the Bogolubov-de Gennes (BdG) equations. With
 phase fluctuations, it
 is only an approximation. 
The average current\cite{Devillard} reads:

\begin{eqnarray}
\langle I \rangle \, =\, {16e \vert \Gamma \vert^4 \over h \hbar W^4}
 \vert \Delta\vert^2 \int_0^{2{eV \over \hbar}}
 P(\Omega) \, d \Omega \, \int_0^{{e V \over \hbar} - {\Omega \over 2}}
D^{-1}(\omega) \, d \omega,
\end{eqnarray}
with
\begin{eqnarray}
D(\omega) \, = \, 
\Biggl({\vert \Delta \vert^2 \over \hbar^2} - \omega^2\Biggr)
 \Biggl(1 + {\vert \Gamma\vert^4 \over W^4}\Biggr)^2
 + 4 {\vert \Gamma \vert^4 \over W^4} \omega^2,
\end{eqnarray}
$W$ being the bandwidth and $\Gamma$ a hopping parameter.
 For tunnel contacts, $\vert \Gamma \vert \ll W$. A perfect contact corresponds
 to $\Gamma = W$ exactly. See
 Ref. \onlinecite{MartinRodero} for details.

We would like to compare two cases, with and without LRI. For JJ chains, LRI 
are usually not taken into account.
 One exception are the models of JJ chain considered in Ref. 
\onlinecite{MSChoi}, where long range 
interactions in the capacitance matrix were examined. These models
 yield physical results which are esentially the 
same as the usual XY model, except in the case where capacitance 
of the islands to the ground, $C_g^{-1}$ is strictly zero. Only in this case
 does the BKT transition disappear and the chain is always in the disordered
 phase, because of the proliferation of unbound vortices. 

In most
 experimental situations, 
it does not seem to be realistic to assume that $C_g^{-1}$
 is really completely 
negligible with respect to nearest-neighbor islands capacitive coupling
 (or even with respect to longer range inter-island capacitive couplings). One 
exception may be the case of the experiments of Ref.
 \onlinecite{GuichardBuisson}. By contrast, the model 
studied in this paper always keeps a non-zero capacitance
 to the ground and still always gives unbound vortices.

Now, we would like to compare the results of our calculation with a model commonly used
 in the literature. In this latter model, a nearest-neighbor interaction
 bewteen island charges is included, and there is no LRI. We discuss qualitatively
 this case and compare the predictions from such a model with our case. 
A term of the form
 $\sum_i q_i q_{i+1}/(2C_1)$ must be added to the Hamiltonian,
 see Eq.(\ref{HJJA}). Qualitatively, this
 will add some charge fluctuations and should
 decrease the phase correlations.
 In a more quantitative way, this will change for example the correlator
 of $\theta$ in Fourier space Eq. (\ref{thetaqw}) to
\begin{eqnarray}
\langle \theta_{q,\omega} \theta_{-q,- \omega} \rangle \, =\, 
\,
{\pi K \over q^2 \bigl\lbrack 1 + {C \over C_1} \cos(qa) \bigr\rbrack + \omega^2},
\end{eqnarray}
where $a$ is the spacing between Josephson junctions. 
Calculations are straightforward and the main changes with respect to
 the case with just a local Coulomb term are the following. 
In the quasi long range phase, for frequencies much smaller than $\omega_p$,  
the plasma frequency is now $\omega^{\prime}_p = \omega_p \sqrt{1 + {C \over C_1}}$,
 and thus the constant $K$ has to be changed accordingly 
to $K^{\prime} = K \sqrt{1 + {C \over C_1}}$. This no longer holds 
for frequencies of order $\omega_p$. The critical point will be shifted. 
Defining $t \equiv \hbar \omega_p/J$, the usual BKT transition which occurred
 at $t=t_c={\pi/2}$ will now take place
 at $t^{\prime}_c = {\pi \over 2 \sqrt{1 + {C \over C_1}}}$.  
In the disordered phase, but still no too far from criticality, the correlation
 length will be depressed. However, the phase correlator will retain approximately
 its Ornstein-Zernike form, Eq. (\ref{OZnormal}), with different values of 
$\omega_p$ and $\xi$. The important thing is that the pre-exponential factor
 is still in $1/\sqrt{\vert {\bf r} - {\bf r}^{\prime}\vert}$.

As an example, we take a perfectly transmitting interface and a situation where
 the Josephson chain would be insulating (disordered  phase) even in absence of LRI. 
The presence of LRI will modify the correlation length $\xi$ but also 
alter the form of the correlator $P(\Omega)$. 
Then, 
$D(\omega)$ is a constant and the current assumes the form
\begin{eqnarray}
\langle I \rangle \, = \, p_0 \, 
\Bigl({e \omega_p \over \pi}\Bigr)
 \int_0^{X} 
{\exp\Bigl(- B  \vert \ln \, x \vert^{3/2}\Bigr) \over \sqrt{x}} \,
(X-x) \, dx,
\end{eqnarray}
with 
\begin{eqnarray}
X \, =\, {2 eV \over \hbar \omega_p} - \xi^{-1}.
\end{eqnarray}
$\langle I \rangle$ remains zero for voltages smaller than
 $V_0 = {\hbar \omega_p \over 2 e} \xi^{-1}$, just as without LRI. 
The new thing is that, for voltages slightly larger than $V_0$, 
$\langle I \rangle$ no longer behaves as $(V-V_0)^{3/2}$. 
We introduce the dimensionless quantity 
$l^{-1} \, = \, \exp(-B^{-2/3})$ and the corresponding
 voltage $V_1 = {\hbar \omega_p \over 2 e} (l^{-1} + \xi^{-1})$. $V_1$ is a crossover
 value of the voltage, not a sharp threshold as $V_0$. 
For voltages $V$, such that $V_0 < V \ll V_1$, then, $X \ll l^{-1}$ and
 $\langle I \rangle$ behaves as
\begin{eqnarray}
\langle I \rangle  \, = \, 
C_2 \sqrt{{V - V_0 \over V_0}} \,
 \exp\biggl(- B \Bigl\vert \ln \Bigl({V-V_0 \over V_0} \xi^{-1}\Bigr) 
\Bigr\vert^{3/2}
\biggr),
\end{eqnarray} 
with $C_2$ a constant.
Thus, the current starts to increase faster than any power law of $(V-V_0)$. 
If, on the other hand, $V$ is larger than $V_1$, then,
$\langle I \rangle$ increases again as $(V-V_0)^{3/2}$. 
For even larger voltages, properties of the phase correlator at short time scales
 are needed. We expect that, the LRI will not be relevant on these scales and
 that the linear behavior of the $I-V$ curves in the usual case will be retrieved. 
This is illustrated in Fig. 5, where we plot $\langle I \rangle$ in units of
$e \omega_p/\pi$ versus $eV/(2  \hbar \omega_p)$, 
for $t = 2 \pi$.
 Since our calculations describe only moderate LRI, for the LRI parameter 
$A$, we take $A = 0.2$ which results in $B=0.316$. 
For the inverse correlation ``length'' $\xi^{-1}$,
 we take\cite{BerezinskiiKosterlitzThouless} it to
 be $\exp\Bigl(-1.5/\sqrt{t-t_c}\Bigr)$. 
This serves just as a rough estimate since the exponential behavior 
of the correlation length applies only (above and)
 close to the usual BKT transition point. It leads to $\xi^{-1}/4 \simeq 0.125$, 
to be compared to $l^{-1}/4 \simeq 0.029$. We see that the Andreev 
current is strongly depressed by LRI. We also compare with the model
 without LRI but with nearest neighbor Coulomb interactions, mentioned above. 
We have taken 
${C \over C_1} = 0.25$. This causes the voltage threshold $V_0$ to be increased to
 a slightly higher value $V^{\prime}_0$, but the current still starts 
as $(V-V^{\prime}_0)^{3/2}$; see Fig. (5).

To be more quantitative,
 we make a logarithmic plot, on Fig. 6, of $\langle I \rangle$
 versus $\bigl\lbrack (V-V_0/V_0) \bigr\rbrack \xi^{-1}$. 
Whereas the curve without LRI, 
for $A \simeq 0$ has a slope $3/2$, this is not the case
 with LRI. The solid curve catches only with the curve without LRI,
 for $V-V_0$ of the order of $0.1$ to $1$, which is 
slightly larger than of order $l^{-1}/4$. For the model with nearest-neighbor
 interaction, we plot instead $\langle I \rangle$ 
versus $\bigl\lbrack (V-V^{\prime}_0)/V^{\prime}_0\bigr\rbrack \xi^{\prime \,\,-1}$. 
The curve (dash-dotted line) is still very close to a straight line of slope $3/2$.  

In the case where the Josephson chain would be in the (quasi) ordered phase, 
$t < t_c$, without LRI, LRI impose a correlation length $\xi_1$ of order
\begin{eqnarray}
\xi_1 \, = \,
A^{-{1 \over 2 \sqrt{t_c-t}}},
\end{eqnarray}
and thus a threshold voltage $(\hbar \omega_p/2e)\, \xi_1^{-1}$.
\begin{figure}[h] 
\epsfxsize 8. cm  
\centerline{\epsffile{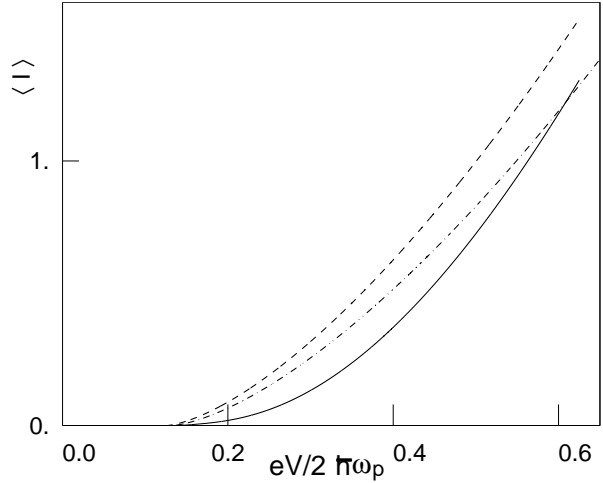}}
\caption{Andreev current $\langle I \rangle$, 
$\lbrack$ in units of $e \omega_p/\pi$ $\rbrack$, 
as a function of applied voltage $V$, 
$\lbrack$ in units of $2 \hbar \omega_p/e$ $\rbrack$,
 for a perfectly transmitting interface and $\hbar \omega_p/J = 6.28$,
 with LRI and $A=0.2$ (solid line), without LRI ($A=0)$ but with a nearest-neighbor interaction
 (dash-dotted line), with only on-site interaction (dashed line).}
\end{figure} 

\begin{figure}[h] 
\epsfxsize 8. cm  
\centerline{\epsffile{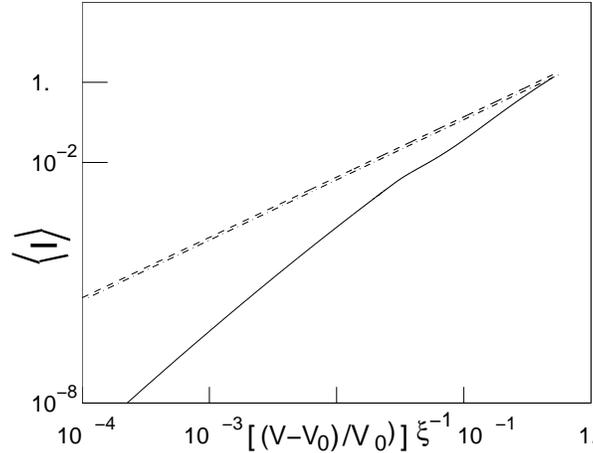}}
\caption{$\log-\log$ plot of Andreev current $\langle I \rangle$, 
as a function of 
$\Bigl\lbrack (V-V_0)/V_0 \Bigr\rbrack \, \xi^{-1}$, (see text), same legend and values as Fig. 5, 
except for the case with a nearest-neigbor interaction
 (dash-dotted line). In this case the threshold value $V^{\prime}_0$ is
 different from the threshold $V_0$ for the two other cases and 
we plot
 $\Bigl\lbrack (V-V^{\prime}_0)/V^{\prime}_0 \Bigr\rbrack \, \xi^{\prime \,\,-1}$ 
 instead on abscissa.}
\end{figure}

\section{Conclusion}
\label{sec:conclusion}

A study of the influence of unscreened long range Coulomb interaction was carried out for 
one-dimensional array of Josephson junctions. This can also have implications for
 one-dimensional thin superconducting wires. We focussed on the case where the initial
 term describing LRI was much smaller than Josephson coupling and 
local Coulomb blockade term. Deriving a Hamiltonian describing 
 the interaction of phase slips, we examined how the KT flow is modified by LRI. 

Implications on Andreev reflexion between a normal metal and a superconductor
 was examined and LRI introduce a new voltage scale. The analytical form of the $I-V$ 
characteristics  was derived for any value of the transparency 
of the normal-superconductor interface, (from tunnel contact to perfect interface).  
If we start from a situation where the charging energy dominates the Josephson coupling,
 corresponding to the JJ chain being insulating, the introduction of LRI changes the 
behaviour above the threshold voltage $V_c$ from a $(V-V_c)^{3/2}$ behavior to
 a sharper increase, faster than any power-law of $(V-V_c)$. This lasts up to
 a crossover voltage $V_1$, whose value depends on the strength of 
the long range Coulomb interactions.
 If, on the contrary,  we start from a situation where the Josephson coupling 
dominates without LRI (quasi ordered phase), 
then LRI will always generate a threshold voltage $V_0$ below
 which no Andreev current can flow. This threshold $V_0$ goes as 
$A^{{1 \over \sqrt{t_c-t}}}$ where $A$ is proportional to the strength of
 the LRI and $t-t_c$ is the distance to the critical point in the absence
 of LRI, with $t = \hbar \omega_p/J$ and $t_c= \pi/2$. 
With LRI, above $V_0$, 
the current starts to increase faster than any power law. 
 The form of the phase correlator
 obtained in this paper
 could be used for calculation of other physical quantities
 relevant to JJ chains or to one-dimensional superconducting thin quantum wires. 

Such calculations are not directly relevant for
 high-$T_c$ superconductors\cite{Efetov}, because of their layered
 structure and their strong anisotropy in the direction
 transverse to the planes, whereas the former 
calculation remained striclty one-dimensional. 

For designed JJ chains, the inverse capacitance is generally cut exponentially with
 a screening length of the order of $\lambda_s \sim a \sqrt{C/C_g}$,
 with usually $\lambda_s \simeq 30 a$, ($a$ is the spacing between neighboring
 junctions), \cite{GrabertDevoretproceedings,Odintsov}.
 The number $N$ of junctions\cite{Chow}
 can be up to $255$, so that $\lambda_s \ll L \equiv N \, a$. Thus, the 
model considered here should not be construed as a
 model for fabricated JJ chains but rather
 an attempt to include LRI, which do exist in isolated thin wires.
 However, we can use the above calculations, replacing the badly screened
 Coulomb potential by the exponentially screened potential $U_{ij}$,
 see Eq. (\ref{Uij}). 
This leads qualitatively to an interaction between vortices 
$V^{\prime}_{\rho,\rho^{\prime}}$ which has the following properties.
For distances $\rho$ much smaller than $\lambda a$, 
$V_{\rho,\rho^{\prime}} = {\cal D} \sqrt{{J \over 2 E_0}} \lambda^{-1/2} (a/\rho)\,\ln \rho$,
 with ${\cal D}$ a constant of the order of ten. 
The strength of the
 interaction between vortices decays very rapidly, (much faster than for
 the badly screened Coulomb interaction). 
This lasts till $\rho$ becomes of the order of
 $\lambda a$, where $V^{\prime}_{\rho,\rho^{\prime}}$ stops to decrease and 
behaves as
 $V_{\rho,\rho^{\prime}} = {\cal D} \sqrt{{J \over 2 E_0}} \lambda^{-3/2} \, Ln \rho$, 
for large distances, i.e. $\rho \gg a \lambda$,

Theoretically, the RG group is started at coarse graining
 scales smaller than $\lambda a$ and 
 as the RG flow proceeds, the interaction 
between vortices gets smaller. This may tend to drive towards 
the disordered region. This is not unexpected, given the large amount of Coulomb
 repulsion provided by the charge coupling of the islands which are less
 than $\lambda a $ apart. When RG proceeds further and 
the coarse-graining length becomes larger than $\lambda a$, then,
 the interaction 
$V_{\rho,\rho^{\prime}}$ bewteen vortices ceases to decay and becomes
 logarithmic with distance. 
According to the value of the coupling reached at that moment, we can flow either
 to the disorded region or to a quasi-ordered fixed point. The screening of the Coulomb
 interaction renders quasi long range-ordering more favorable. 
In particular, contrary to the badly screened interaction case, 
phase correlations do not always decay exponentially at the end. 

Let us suppose first that $J$ is sufficiently large so that we end in the (quasi)-ordered 
phase, $P(\Omega)$ will decay as a power law for small $\Omega$, 
 with some exponent $\alpha$, for $\Omega$ smaller
 than $\omega_p/\lambda$. For $\Omega$ larger than $\omega_p/\lambda$, the decay of $P(\Omega)$ 
corresponds to a larger effective vortex interaction,
 and phase correlations are enhanced, leading to a larger $P(\Omega)$. 
As we go to large distances (or longer times or smaller voltages),
 vortex interactions get reduced until the distance $\lambda a$ is reached.
 This is due large Coulomb repulsion at shorter lengthscales. This has experimental consequences on 
the shape of $I$ vs. $V$ curves.
 Frequencies smaller than $\omega_p/\lambda$ correspond to voltages 
smaller than $\hbar \omega_p/e$. For these voltages, $I-V$ curves
 will start as power law with some exponent,
 normally as $V^{\alpha}$ with
 $\alpha= {1 + {1 \over \pi}\sqrt{{E^{\prime}_0 \over J^{\prime}}})}$,
 where $E^{\prime}_0$ and $J^{\prime}$ are renormalized values of $E_0$ and $J$. 
When $V$ becomes larger than $\hbar \omega_p/e$, the current is enhanced
 and increases faster than $V^{\alpha}$. This remains true for
 voltages still smaller than $\hbar \omega_p/e$. 

Now, let us suppose that $J$ is not so large and 
one ends in the disordered phase. There is a threshold voltage
 but beyond $V_c$, the current starts to increase as $(V-V_c)^{3/2}$ in 
 the usual theory. Inclusion of screened interaction leads to an increase 
which is faster than $(V-V_c)^{3/2}$, which could also be tested experimentally. 

The model with badly screened interactions may also have applications 
 for very long (but not unrealistically long) 1d superconductor thin wires. 
More precisely in thin wires (typically $3$ to $5$ ${\rm nm}$ diameter),
 the interaction between phase slips has the same structure as the Hamiltonian
 (1), but the physical origin is somewhat different \cite{Zaikinreview}.
First, in concrete situations, it is not the two-dimensional BTK transition 
which is seen but rather a one-dimensional Schmid transition\cite{Schmid}.
 We briefly recall the reasons below.

 In experiments with very homogenous thin wires in the difusive limit,
 of diameter 3 to 10 nm, such as the ones used in Refs. \onlinecite{Tinkhamwires,Chang}
 the maximum length $L$ is  about $1 {\rm \mu} {\rm m}$.
 Operating temperatures are between $100 {\rm mK}$ and a few Kelvin. 
If the length were infinite, renormalization had to be stopped
 when coarse graining length $\tau$ on the imaginary time axis reaches 
 $\beta \hbar$. However, in practise, two-dimensional renormalization
 has to be stopped when it reaches $L/v_{MS}$ 
where $v_{MS}$ is the Mooj-Sch{\"o}n velocity\cite{MooijSchon}. 
$v_{MS} \sim \omega_p d/C_L$ is equivalent of the spin-wave velocity,
 where $C_L$ is the capacity per unit length and $d$ the diameter of the wire.
$v_{MS}$ is of the order $10^{5} m. s^{-1}$, so that
 for a $1 {\rm \mu} {\rm  m}$ long wire, $L/v_s$ is of the order $10^{-11}{\rm s}$ and for temperatures
 lower than $0.75 {\rm K}$, renormalization has to be stopped first on the 
 $x$ axis, when the coarse graining length reaches the size of the wire. 
We end up with charges in 1 dimension 
 (as for a single Josephson junction) which interact logarithmically 
as ${\cal I}_{ren} Ln \vert r_i - r_j \vert$, with
 ${\cal I}_{ren}$ a renormalized coupling. Then, renormalization proceeds solely
 on the imaginary time direction. Depending on the value of ${\cal I}_{ren}$, 
fugacity of vortices renormalizes to zero or not, resulting in the usual
 Schmid transition. The spatial structure
 on the x axis has been completely lost and the long-range interaction 
 do not play any role, except that they changed
 slightly the values of ${\cal I}_{ren}$. 
Thus, in this case, LRI do not seem to play an important role. Thus, in order to see
 the influence of LRI, longer wires are needed.

For longer wires (but not totally unrealistic,
 a few microns long), we can expect temperatures larger than $\hbar v_s/(L k_B)$,
 so that renormalization has to be cut on the imaginary time axis first. Then, LRI 
will drive the resistance $R$ to a finite value. If $R$ versus $T$ curves are plotted,
 no curve should plunge towards zero resistance, unless $T$ becomes
 of the order of $\hbar v_s/(L k_B)$, which is about $0.5 \, {\rm K}$ for
 $L=10 \mu {\rm m}$. 

There are difficulties to see LRI. 
One could be that stray capacities
 (unavoidable in experiments) can also cut the long-range term in
     the capacitance for small $q$ (in real space) so that
    $C^{-1}(k)$ is no longer of the form $\ln k$ for small $k$. This
 depends on the charateristics of the wires. 
Another problem is that the action for a single phase slip, which controls its bare fugacity, is
 strongly dependent on temperature and on the diameter of the wire. This dependence
 seems to be more important than the interaction bewteen phase-slips in determining
 the shape of the curves showing the resistance $R$ versus $T$, for different wire thicknesses
\cite{Zaikinreview}.
  
  Long range interactions are originally not screened
  in isolated one-dimensional wires. However, the presence of an environment could
 alter the screening. Situations where a plate or a wire is placed in the vicinity
 of a superconducting thin wire are now experimentally\cite{Snyder}
and theoretically studied \cite{RefaelDemlerOregFisher,LobosGiamarchi}. This aims   
at providing mechanisms of dissipation
 so as to damp the phase fluctuations. Though we did not
 consider  in this paper coupling to an external environment,
 our calculation 
may be viewed as a first step towards understanding the role of LRI, including, both spin waves and
 phase slips (solitons). 

\acknowledgments

We thank A. Cr\'epieux and J. Rech for reading of the manuscript.

\appendix

\section{}

In this appendix, 
we derive the RG flow in the presence of long-range interactions. We define 
$g$ as 
\begin{eqnarray}
g(R_{i,j})\,\equiv \, 
{\cal J} \, {\cal F}({\bf r}_i-{\bf r}_j),
\end{eqnarray}
with
\begin{eqnarray}
R_{i,j} \, = \, \vert {\bf r}_i - {\bf r}_j \vert.
\end{eqnarray}
Integration has to be done over ${\bf r}_j$ in the whole plane except the two circles of
 radii $\tau$ around ${\bf r}_k$ and ${\bf r}_l$. 
Setting ${\bf r} \, = \, ({\bf r}_l + {\bf r}_k)/2$ and
 denoting by ${\cal K}$ the vortex fugacity, the partition function $Z$ reads
\begin{eqnarray}
Z \, &=& \, 
\sum_{n} {1 \over n!^2} \, {\cal K}^{2n}\,
\int_{{\rm D}} d^2{\bf r}_1 \,...\, \int d^2{\bf r}_n 
\Biggl\lbrace
1 \, + \, {{\cal K}^2 \over (n+1)^2} \sum_{i=1}^{n+1} \sum_{j=1}^{n+1}
  2 \pi \tau \, d \tau
\nonumber \\
&\,\,& \int_{{\rm D}} d^2 {\bf r}_j
\Bigl\lbrack 1 \, + \,
\sum_k  g^2(R_{j,k}) {\tau^2 \over \vert{\bf r}_j - {\bf r}_k \vert^2} \, 
+ \sum_{k,l, \, k \not= l} g(R_{j,k}) g(R_{j,l}) q_k q_l \, 
\Bigl({\tau^2  ({\bf r_j}-{\bf r}_k) ({\bf r_j}-{\bf r}_l) \over
\vert{\bf r}_j - {\bf r}_k\vert^2 \, \vert{\bf r}_j - {\bf r}_l 
\vert^2}\Bigr)
\Bigr\rbrack
\Biggr\rbrace \nonumber \\
&\,\,& \exp \Bigl\lbrace 
\sum_{i,j} {\cal F}({\bf r}_i-{\bf r}_j) \, q_i q_j \,
 \ln\vert {\bf r}_i-{\bf r}_j\vert,
\Bigr\rbrace.
\end{eqnarray}
where ${\rm D}$ is the whole plane except aera inside the circle
 of radii $\tau$ around the ${\bf r}_i$, $i=1,n$.
Let us denote by
\begin{eqnarray}
{\cal I} \, =  \int_{{\cal D}}
g(R_{j,k}) \, g(R_{j,l}) {({\bf r}_j - {\bf r}_k).({\bf r}_j + {\bf r}_l) \over 
\vert {\bf r}_j - {\bf r}_k \vert^2 \, \vert {\bf r}_j + {\bf r}_l \vert^2}\,
d^2{\bf r}_j,
\end{eqnarray}
where ${\cal D}$ is the whole plane except the two circles
 of radii $\tau$ around ${\bf r}_k$ and ${\bf r}_l$.
Given the value of ${\cal F}(r)$, see Eq. (\ref{fcaleq}), we have
\begin{eqnarray}
{\cal I} \, \simeq \, 2 \pi \, \ln(R_{k,l}/\tau) \,
 \Bigl\lbrack 1 - {4 \over 3} {\cal A} \sqrt{\ln(R_{k,l}/\tau)}\,\Bigr\rbrack,
\end{eqnarray}
for ${\cal A} \ll 1$.
Denote by
\begin{eqnarray}
{\cal C} \, = \,
2 \pi \tau \, d \tau \, 
\biggl\lbrack S \, - \,
\sum_{k,l \,k \not= l} 2 \pi \tau^2 {\cal J}^2
\Bigl\lbrace 1 - {4 \over 3} {\cal A} \sqrt{\ln(R_{k,l}/\tau)} \Bigr\rbrace
q_k q_l \, \ln\Bigl\vert{{\bf r}_k - {\bf r}_l \over \tau}\Bigr\vert\,
\biggr\rbrack,
\end{eqnarray}
where $S$ is the aera.
Collecting all the terms, 
\begin{eqnarray}
Z \, = \, \sum_n {1 \over n!^2} \, {\cal K}^{2n}
\int d^2{\bf r}_1 ... \int d^2{\bf r}_n 
\biggl\lbrack 1 + {{\cal K}^2 \over (n+1)^2} 
\sum_{i=1}^{n+1} \sum_{j=1}^{n+1} {\cal C} \biggr\rbrack
\exp\biggl( \sum_{i,j}
g(R_{i,j})
 \, q_i q_j
\, \ln \Bigl\vert{{\bf r}_i - {\bf r}_j \over \tau}\Bigr\vert \biggr).
\end{eqnarray}
Using $e^x = 1+x$, for small $x$,
\begin{eqnarray}
Z \, &=& \!\!\sum_n  {1 \over n!^2}\, {\cal K}^{2n} \, 
\Biggl(\exp\Bigl\lbrace \sum_{i=1}^{n+1} \sum_{j=1}^{n+1}
 {{\cal K}^2 \over (n+1)^2}
 2 \pi \tau \, d \tau \, S \Bigr\rbrace \Biggr)\, \nonumber \\
&\,\,& 
\int d^2{\bf r}_1 ... \int d^2{\bf r}_n 
\exp\Biggl\lbrace
\sum_{i,j} g(R_{i,j}) q_i q_j \ln\Bigl({R_{i,j} \over \tau}\Bigr)
 \nonumber \\
&\,&
- (2 \pi)^2 \tau^4 \, {\cal K}^2 \, {d \tau \over \tau} \, {\cal J}^2
\sum_{k,l \, k \not= l}
 q_k q_l \, \ln \Bigl\vert {R_{k,l} \over \tau} \Bigr\vert
\Bigl(1 - {4 \over 3} {\cal A} \sqrt{\ln(R_{k,l}/\tau)}\,\Bigr) \Biggr\rbrace.
\end{eqnarray}

When changing $\tau$ into $\tau + \, d \tau$, this leads only to the 
renormalization of the fugacity ${\cal K}$ into
 ${\cal K} \Bigl(1 -  {\cal J} {d \tau \over \tau}\Bigr)$ 
replacing ${\cal F}(r_i - r_j)$ by its initial value, which is one.

After changing $(k,l) \rightarrow(i,j)$ in the second summation,
the part in the exponential reads
\begin{eqnarray}
\sum_{i,j} q_i q_j 
\biggl\lbrack {\cal J} {\cal F}(R_{i,j}) \, \ln\Bigl({R_{i,j}  \over \tau}\Bigr) 
 \, - (2 \pi)^2 \tau^4 {\cal J}^2 \, {\cal K}^2 \Bigl({d \tau \over \tau}\Bigr) \,
\, \ln\Bigl({R_{i,j}  \over \tau}\Bigr) 
\Bigl(1 - {4 \over 3} {\cal A} \sqrt{\ln(R_{i,j}/\tau)}\Bigr) \biggr\rbrack.
\end{eqnarray}

If ${\cal F}(R_{i,j})$ were a constant, we would retrieve the usual KT flow. However,
 since we are looking at larger and larger lengthscales, the effective coupling will not
 only decrease because of RG but also because of its $\sqrt{\ln R}$ dependence. 
The smallest $R$ contributing to $Z$ is $\tau$. 
Thus, adding  a term proportional to $- {4 \over 3} {\cal A} \sqrt{\ln(\tau)}$ 
will take this into account. 

Now, the partition function has to be rearranged so as to look as 
its original form,
 with renormalized parameters and an overall multiplicative factor,
 setting $z = \ln \tau$, we obtain
\begin{eqnarray}
{d{\cal J} \over d z} \, &=&\, - 4 \pi^2 {\cal J}^2 {\cal K}^2 \tau^4 - 
{4 \over 3} {\cal A} \sqrt{\ln \, \tau}, \\
{d {\cal K} \over d z} \, &=& \, - {\cal J} \, {\cal K}.
\end{eqnarray}
Switching to the usual variables
\begin{eqnarray}
X \, &=& \, {\cal J} - 2, \\
Y \, &=& \,\bigl(4 \pi {\cal K} \tau^2\bigr)^2, \\
\end{eqnarray}
and linearizing around $X=0$, the flow reads
\begin{eqnarray}
{d X \over dz} \, &=& \, -Y - {\cal B} \sqrt{z}, \\
{d Y \over dz} \, &=& \, -2 XY,
\end{eqnarray}
with ${\cal B} = 4 \sqrt{A}/3$.

\section{}

Here, 
we give details on the contributions to the vortex part of the correlation function
 $g_V({\bf r} - {\bf r}^{\prime})$ coming from smooth variations of 
 $v({\bf r})$. 
The first term to evaluate is 
\begin{eqnarray}
T^{\prime}_1 \, = \, 
-  \int \int^{\prime} \Delta v({\bf R}) \, d^2{\bf R},
\end{eqnarray}
where the double integral extends over the whole shaded
 surface in Fig. 4.
 $\Delta$ is the Laplacian operator (with respect to variable
 ${\bf R}$). In this calculation, 
${\bf r^{\prime}}$ and ${\bf r}$ are fixed, and, without restriction, 
${\bf r}$ is taken to be zero for simplicity. If $R$ is the modulus of
 ${\bf R}$, since $v$ decreases as $\ln(R)^{-1/2}$ for large $R$, it turns out that 
$\Delta v$ behaves, as $R^{-2} \, (\ln \,R)^{-3/2}$. Thus, 
$v \, \Delta v$ behaves as $R^{-2} \,(\ln R)^{-2}$.
When surface integration is performed, the integral tends to a constant,
 for $r^{\prime} \rightarrow \infty$ because of the $ (\ln R)^{-2}$ factor.
When exponentiated this will give a constant. Since, we shall neglect
 all terms that depend on $r^{\prime}$, weaker than any power law,
 we shall neglect this factor.

The second term is a line integral over the outer contour 
$\Gamma_{out}$
on Fig. 4, 
\begin{eqnarray}
T^{\prime}_2 \, = \, \int_{\Gamma_{out}} v \, (\nabla v). d {\bf \sigma},
\end{eqnarray}
where $d \sigma$ is the normal to the contour 
$\Gamma_{out}$ which is here chosen as a circle 
of radius ${\cal R}$. 
For all ${\bf R}$'s belonging to $\Gamma_{out}$, $R$ are equal to ${\cal R}$
 and $v$ are of order $(\ln\,{\cal R})^{-1/2}$, the leading terms in  
$(\nabla v)$ are of order ${\cal R}^{-1} (\ln \,{\cal R})^{-1/2}$. 
Thus $T_2$ is at most of order $(\ln \,{\cal R})^{-1}$ and vanishes as
 ${\cal R}$ goes to infinity. 

The third term to evaluate is a line integral over the inner contour $\Gamma_{in}$. 
on Fig. 4.
\begin{eqnarray}
T^{\prime}_3 \, =\, \int_{\Gamma_{in}} v \, (\nabla v). d {\bf \sigma}.
\end{eqnarray}
A calculation completely analogous to the case without LRI shows 
that the leading terms are of order 
 $\int_b^{r^{\prime}}
 {x^{\prime}  \over x^{\prime \,2}   + y^{\prime \, 2}}
\Bigl\lbrace \ln(x^{\prime \,2} + y^{\prime \, 2} ) \Bigr\rbrace^{-1} \, 
dx^{\prime} + \bigl(x^{\prime} \leftrightarrow (r^{\prime} - x^{\prime})\bigr)$,
 with $r^{\prime}$ the modulus of ${\bf r} - {\bf r}^{\prime}$.
 This behaves, for large $r^{\prime}$ as $\ln(\ln \,r^{\prime})$.

\end{document}